\documentclass[aps,prc,preprint,showpacs,amsmath,amssymbd,superscriptaddress,showkeys]{revtex4-1}
\usepackage{graphicx}	
\usepackage{xspace}	
\usepackage{color}
\usepackage{hyperref}
\def \be  {\begin{equation}}
\def \ee  {\end{equation}}
\def \ee  {\end{equation}}
\def \bea {\begin{eqnarray}}
\def \eea {\end{eqnarray}}

\newcommand{\nn}{\nonumber}

\begin{document}
\title{Influence of finite volume and magnetic field effects on the QCD phase diagram}

\author{Niseem~Magdy} 
\email{niseem.abdelrahman@stonybrook.edu}
\affiliation{Department of Chemistry, State University of New York, Stony Brook, New York 11794, USA}

\author{M.Csan\'ad} 
\email{csanad@elte.hu}
\affiliation{Department of Chemistry, State University of New York, Stony Brook, New York 11794, USA}
\affiliation{E\"otv\"os University, Department of Atomic Physics, H-1117 Budapest, Hungary}

\author{Roy.A.Lacey} 
\email{roy.lacey@stonybrook.edu}
\affiliation{Department of Chemistry, State University of New York, Stony Brook, New York 11794, USA}

\begin{abstract}
The $2+1$ $SU(3)$ Polyakov linear sigma model (PLSM) is used to investigate the respective influence of a finite volume and a magnetic field on the quark-hadron phase boundary in the plane of baryon chemical potential  ($\mu_{B}$) vs. temperature ($T$) of the QCD phase diagram. The calculated results indicate sizable shifts of the quark-hadron phase boundary to lower values of  $(\mu_{B}~\text{and}~T)$ for increasing magnetic field strength, and  an opposite shift to higher values of $(\mu_{B}~\text{and}~T)$ for decreasing system volume. Such shifts could have important implications for extraction of  the thermodynamic properties of the QCD phase diagram from heavy ion data.
\end{abstract}


\maketitle
\makeatletter
\let\toc@pre\relax
\let\toc@post\relax                 
\makeatother 
%
%
\section{Introduction}
A major impetus for current heavy ion research is the prospect of obtaining profound insights on the 
rich phase structure of strongly interacting matter at high temperature and non-zero net
baryon number density. Ongoing programs at RHIC \cite{RHIC}, the SPS \cite{SPS} and the LHC \cite{ALICE}, as well
as future facilities at FAIR \cite{FAIR} and NICA \cite{NICA} are at the forefront of experimental  efforts designed 
to map the thermodynamic and transport properties of  this strongly interacting QCD matter.
Lattice QCD simulations suggests a smooth cross over phase transition from 
hadronic matter to the quark gluon phase at low density and high temperature \cite{nature05120_Aoki, PRL65_2491_Brown}. 
At high density and low temperature a first order phase transition is 
expected \cite{PRD78_074507_Ejiri, PRD29_Pisarski, NPA504_Asakawa, PRD58_096007_Halasz, PRD67_014028_Hatta, 
PRC79_Bowman}. Both transition domains, crossover and the first-order phase-transition, are connected by the expected 
critical endpoint (CEP), at which the phase transition is likely second order. The beam energy scan (BES) program 
at RHIC, have begun to show striking non-monotoic signatures which could be an indication that the CEP is located 
at high temperature and modest values of baryon chemical potential ($\mu_B$) \cite{Roy_CEP}.

A preponderance of the theoretical studies assume an infinite volume devoid of magnetic fields, for 
the QCD matter produced in heavy ion collisions.  This is in stark contrast to the finite volumes and 
sizable magnetic fields produced in these collisions (both depend on the size of the colliding nuclei, 
the center of mass energy ($\sqrt{s_{NN}}$) and the collision centrality). Therefore, it is important 
to ask whether the combined influence of a finite volume and a strong magnetic field leads to a modification 
of the apparent thermodynamic properties of the produced QCD medium.

The influence of a finite-volume and the presence of a strong magnetic field ($B$) has been widely discussed in the 
literature~\cite{review,fss,fss1,fss2,fss3,Fraga_11,Bhattacharyya,fragamit,G_Endrodi,Klevansky1992,Shushpanov1997,Agasian2000,
Cohen2007,Kabat2002,Marco2012,TNMF14,boom1,boom2,rg1,rg2,rg3,Gatto,gattopnjl,latticemassimo}. 
This includes the effects on the value of the critical temperature, the location of the critical end point and other thermodynamic properties.
Initial studies of the magnetic field effect include lattice QCD (lQCD) studies~\cite{latticemassimo}, the MIT bag model~\cite{fragamit}, 
the Nambu-Jona Lasinio (NJL) model~\cite{boom1,boom2} and  the Linear Sigma Model  (LSM), or Quark Meson model (QM)~\cite{rg1,rg2,rg3},
as well as extensions of the NJL and the LSM involving the Polyakov loop (PNJL and PLSM)~\cite{Gatto,gattopnjl}. The results from these studies, which 
indicate an increase of  the transition temperature $T_c$ with increasing magnetic field~\cite{review},  contrast with the results from 
recent lQCD calculations (with a physical  pion mass $m_{\pi}=140$ MeV) which indicate that  $T_c$ decreases with increasing magnetic field. 
The PLSM and PNJL models have been recently used to study the latter trend for vanishing chemical potential~\cite{costa,TNMF14}. 
         The effects of a finite volume~\cite{fss,fss3} which include a strong influence on the value of the transition temperature ($T_c$), 
the location of the critical end point and other thermodynamic properties,  have been extensively studied with the NJL~\cite{fss1}, 
LSM~\cite{fss2,Fraga_11} and PNJL~\cite{Bhattacharyya} models. However, strikingly different trends for the influence 
of  finite-size effects on $T_c$ have been reported  for the LSM and PNJL models.


 In this work, we use the $2+1$ $SU(3)$ Polyakov Linear Sigma Model (PLSM)~\cite{Tawfik:2014uka,Schaefer:2008ax,Mao:2010} with the 
Fukushima Polyakov loop effective potential~\cite{PNJLsus2p1f1} to investigate the combined effects of finite volumes and magnetic fields at temperatures and chemical potentials akin to those produced in heavy ion collisions over a broad range of beam collision energies. The present work is organized as follows. In section \ref{sec:approaches} we give a brief overview of the PLSM~\cite{Tawfik:2014uka,TNMF14}, as well as the parameters of the model employed in this study. We then present the results of our study on the influence of  finite-volume and the magnetic field  effects on the PLSM order parameters (chiral condensates $\sigma_x$ and $\sigma_y$ and Polyakov loops  $\phi$ and $\phi^*$), thermodynamic properties and  the chiral phase transition in section \ref{sec:III}. 
We conclude with a summary and an outlook in section \ref{sec:conclusion}.

%
\section{The Polyakov Linear Sigma Model (PLSM)}
 \label{sec:approaches}

The $SU(3)$ Linear Sigma Model with $N_f = 2+1$ flavor quarks, can be coupled to Polyakov loop dynamics to formulate
the PLSM \cite{Schaefer:2008ax,Mao:2010,Tawfik:2014uka}.  The associated  Lagrangian is given as;
\begin{eqnarray}
\mathcal{L}=\mathcal{L}_{\rm chiral}-\mathbf{\mathcal{U}}(\phi, \phi^*, T), \label{plsm}
\end{eqnarray}
where the chiral part [quark $q$ and meson $m$ ] of the Lagrangian, $\mathcal{L}_{\rm chiral}=\mathcal{L}_q+\mathcal{L}_m$, has $SU(3)_{L}\times SU(3)_{R}$ 
symmetry \cite{Lenaghan,Schaefer:2008hk}; $\mathbf{\mathcal{U}}(\phi, \phi^*, T)$ represents the Polyakov loop effective potential\cite{PNJLsus2p1f1}. 
This effective potential leads to reasonable agreement with recent lQCD results~\cite{QCDL}.
Other Polyakov loop potentials~\cite{Haas:2013qwp,Schaefer_2009} were also considered. However, the particular choice made for this work does not influence the main conclusions of our work.
\begin{widetext}
\begin{eqnarray}
\mathcal{U}(\phi, \phi^*, T) &=& - b T \left\{54\, \exp\left(-\frac{a}{T}\right) \phi \phi^* + \ln\left[1- 6 (\phi \phi^*)^{2} - 3(\phi \phi^* )^{3} + 4 (\phi^{3} + \phi^{*3})\right] \right\}
\label{Uloop},
\end{eqnarray}
\end{widetext}
with $a=0.664~$GeV and $b=0.0075~$GeV$^3$ \cite{PNJLsus2p1f1}.
The mean field approximation is used to obtain the grand potential \cite{Tawfik:2014uka} as; 
\begin{widetext}
\begin{eqnarray}
\Omega(T, eB, \mu_f) &=& U(\sigma_x, \sigma_y)+\mathbf{\mathcal{U}}(\phi, \phi^*, T) + \Omega_{\bar{\psi}\psi} (T,\mu_f;\phi,\phi^{*},B) 
+ \delta_{0,B}~\Omega_{\bar{\psi}\psi}^{0} (T,\mu_f;\phi,\phi^{*}), 
\label{potential}
\end{eqnarray}
\end{widetext}
where the first term in Eq. (\ref{potential})  is a purely mesonic potential expressed as,
\begin{widetext}
\begin{eqnarray}
U(\sigma_x, \sigma_y) &=& \frac{m^2}{2} (\sigma^2_x+\sigma^2_y)-h_x
\sigma_x-h_y \sigma_y-\frac{c}{2\sqrt{2}} \sigma^2_x \sigma_y \nonumber \\
&+& \frac{\lambda_1}{2} \sigma^2_x \sigma^2_y +\frac{1}{8} (2 \lambda_1
+\lambda_2)\sigma^4_x + \frac{1}{4} (\lambda_1+\lambda_2)\sigma^4_y. \label{Upotio}
\end{eqnarray}
\end{widetext}
Here, $m^2$, $h_x$, $h_y$, $\lambda_1$, $\lambda_2$ and $c$ are model parameters as outlined in Ref.~\cite{Schaefer:2008hk}. 
The values for these parameters,  used in the present study, are tabulated in Table~\ref{par_tab} below. 
Recent studies~\cite{Peter15,Peter16} suggest that better consistency with recent lattice results can be achieved in the $T<T_c$ region, 
by extending the model within the vector meson sector. Such an extension was not included in this work and is not expected 
to strongly affect the qualitative conclusion.

The third term in Eq. (\ref{potential}) $\Omega_{\bar{\psi}\psi} (T,\mu_f;\phi,\phi^{*},B)$ represents the contributions from 
quarks and anti-quarks at a non-vanishing magnetic field strength.  Using Landau quantization and magnetic catalysis 
concepts, this potential can be expressed as~\cite{TNMF14};
%
\begin{eqnarray}
\Omega_{\bar{\psi}\psi} (T,\mu_f;\phi,\phi^{*},B) &=& -2 \sum_{f} \dfrac{|q_{f}| B T}{2 \pi} \sum_{\nu=0}^{\infty} \int \dfrac{d p}{2 \pi} \left(2-1 \delta_{0n}\right)
\\ \nn &&
 \left\{ \ln \left[ 1+3\left(\phi+\phi^* e^{-\frac{(E_{f}^{\nu} - \mu_f)}{T}}\right)\, e^{-\frac{(E_{f}^{\nu} - \mu_f)}{T}}+e^{-3 \frac{(E_{f}^{\nu} - \mu_f)}{T}}\right]\right. 
  \\ \nn &&
  \left.   +\ln \left[ 1+3\left(\phi^*+\phi e^{-\frac{(E_{f}^{\nu} + \mu_f)}{T}}\right)\, e^{-\frac{(E_{f}^{\nu} + \mu_f)}{T}}+e^{-3\frac{(E_{f}^{\nu} + \mu_f)}{T}}\right] \right\},  \label{new-qqpotio}
\end{eqnarray}
%
where $E_{f}^{\nu}$ is the modified quark dispersion \cite{TNMF14} and $\mu_f$ denotes the quark chemical potentials.
The subscript $ f $ runs over different quark flavors. The quark chemical potentials are related to the baryon ($\mu_B$), strange ($\mu_S$) 
and charge ($\mu_Q$) chemical potentials via the following transformations \cite{Wei_13};
\begin{eqnarray}
\mu_{u} &=& \frac{\mu_B}{3} + \frac{2 \mu_Q}{3},\nn\\
\mu_d &=& \frac{\mu_B}{3} - \frac{\mu_Q}{3},\nn\\
\mu_s &=& \frac{\mu_B}{3} - \frac{\mu_Q}{3} - \mu_S,\nn
\label{eq.muqtomuH}
\end{eqnarray}
and $E_{f}^{\nu}$ is given as \cite{TNMF14}:
\begin{eqnarray}
E_{u}^{\nu} &=&\sqrt{p_{z}^{2}~+m_{q}^{2}~+|q_{u}|(2n+1-\sigma) B}, \label{Eu} \\
E_{d}^{\nu} &=&\sqrt{p_{z}^{2}~+m_{q}^{2}~+|q_{d}|(2n+1-\sigma) B}, \label{Ed} \\
E_{s}^{\nu} &=&\sqrt{p_{z}^{2}~+m_{s}^{2}~+|q_{s}|(2n+1-\sigma) B}, \label{Es}
\end{eqnarray} 
where $\sigma$ is related to the spin quantum number $S$ ($\sigma=\pm S/2$). Here,  we have replaced $2n+1-\sigma$ by 
one quantum number $\nu$, where $\nu=0$ is the Lowest Landau Level $m_{f}$, and $f$ runs over the $u$, $d$ and $s$
quark masses, 
\begin{eqnarray}
m_q &=& g \frac{\sigma_x}{2}, \label{qmass} \\
m_s &=& g \frac{\sigma_y}{\sqrt{2}}.  \label{sqmass}
\end{eqnarray} 
%
%
%
\begin{figure*}[tb]
\centering{
\includegraphics[width=4.5cm,angle=-90]{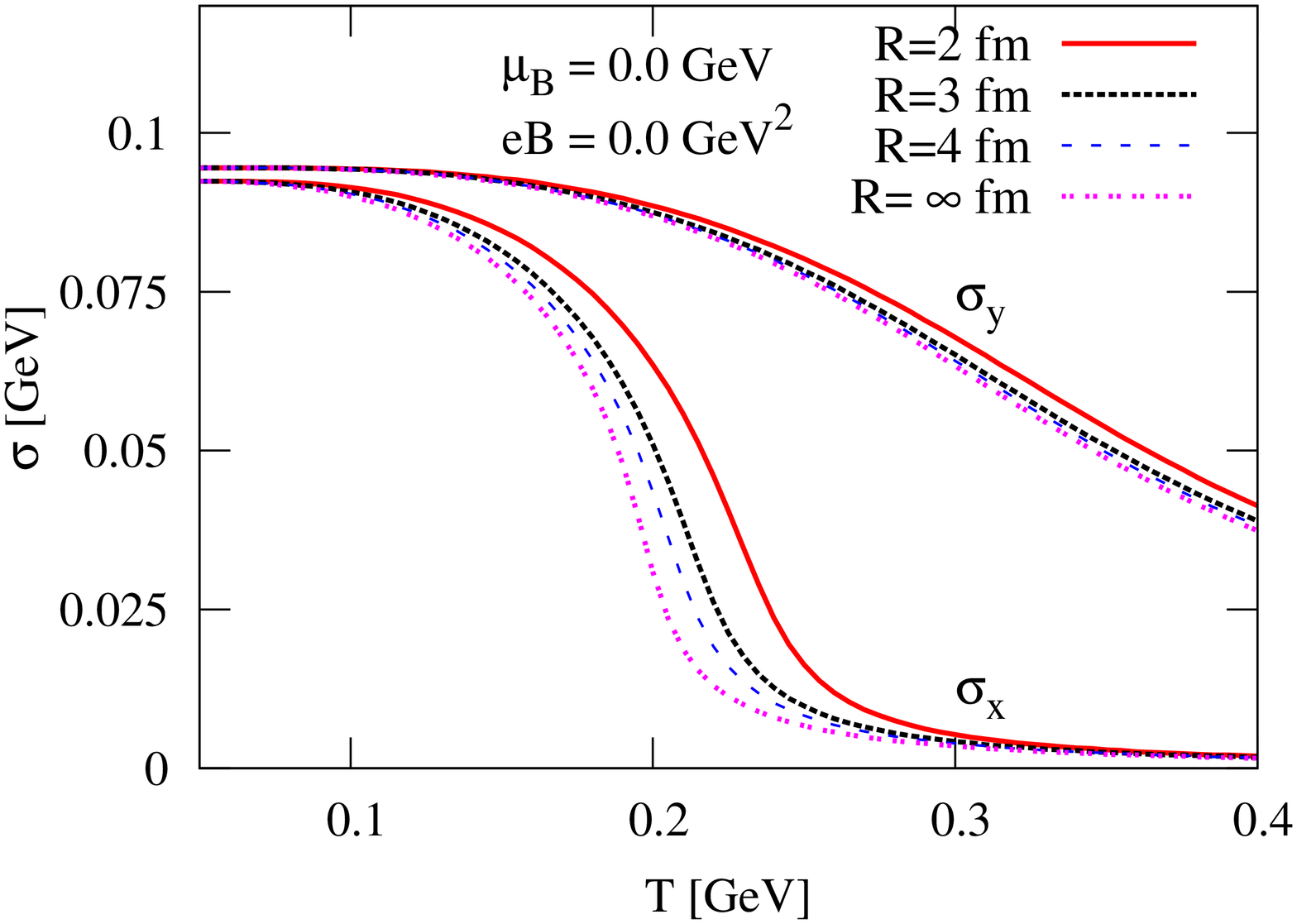}
\includegraphics[width=4.5cm,angle=-90]{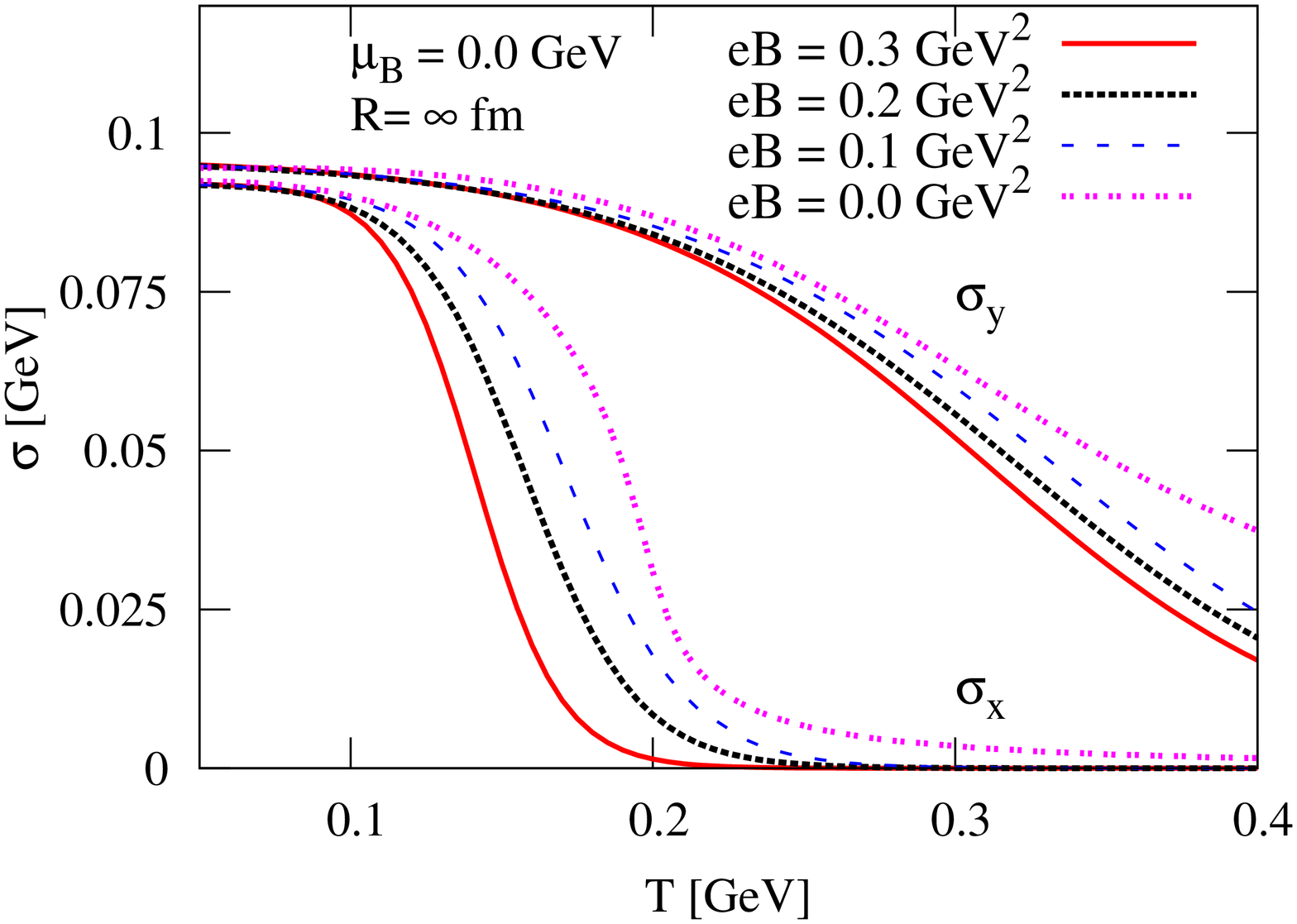}
\includegraphics[width=4.5cm,angle=-90]{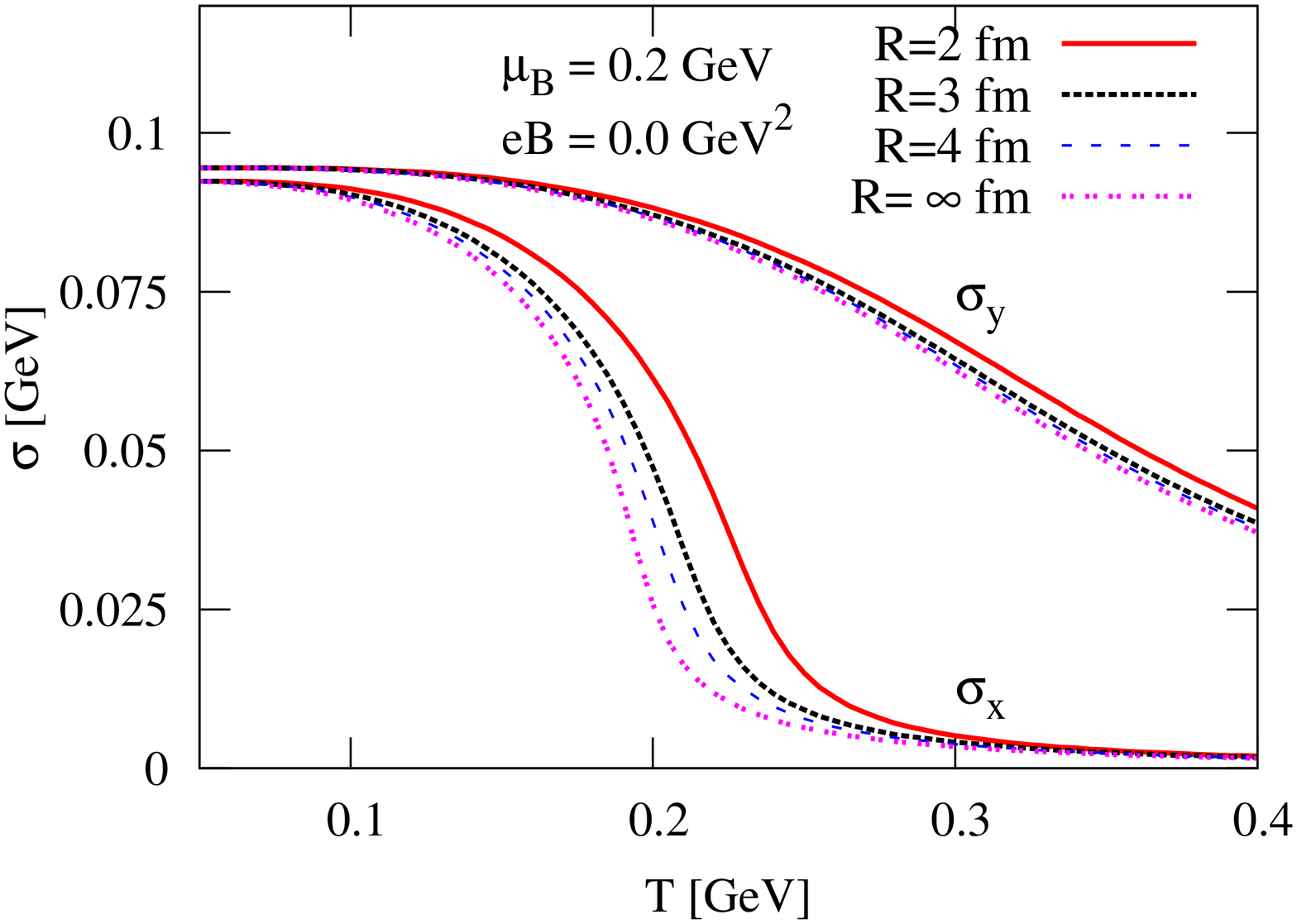}
\includegraphics[width=4.5cm,angle=-90]{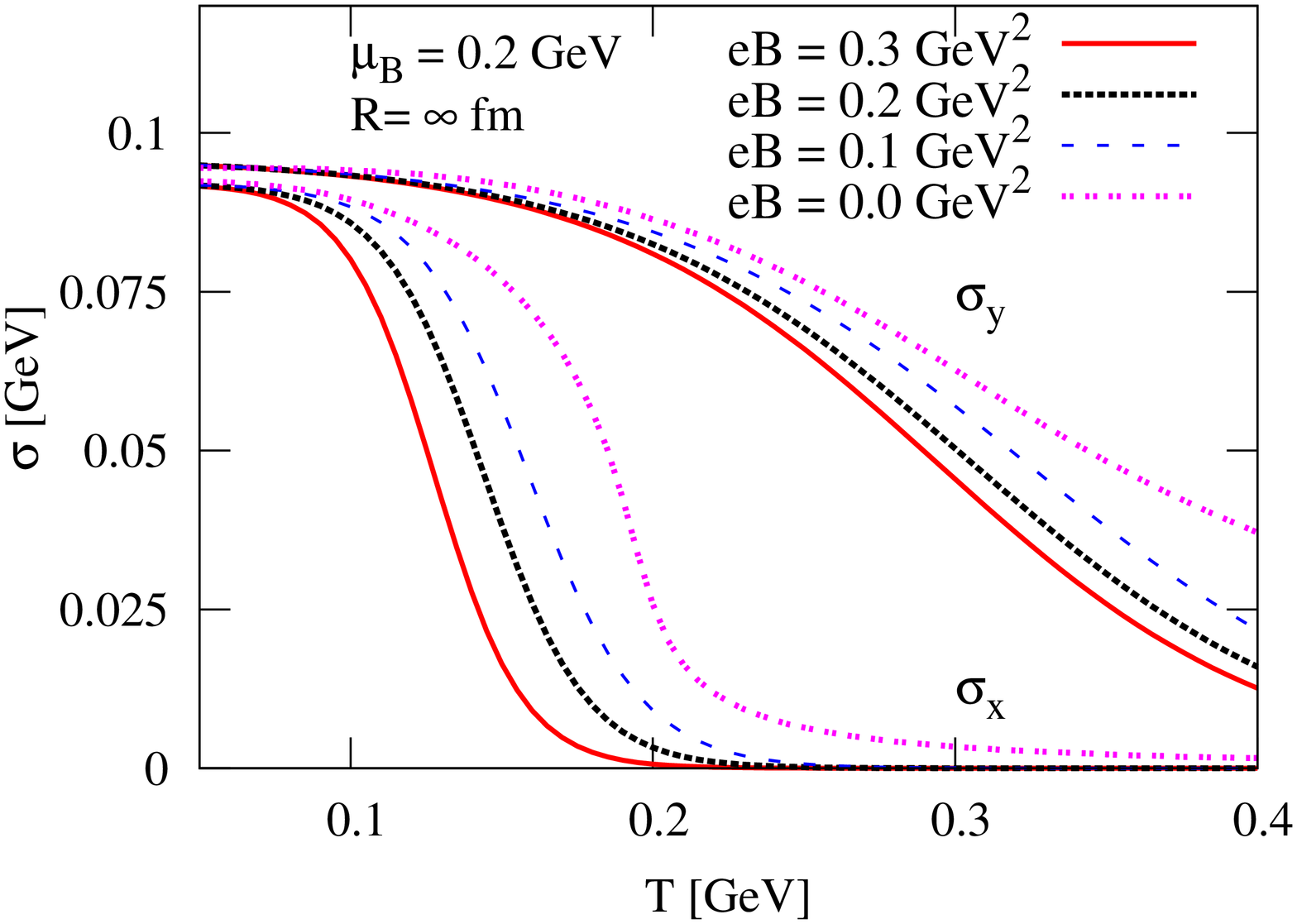}
\caption{(Color online)  Temperature dependence of the chiral condensates $\sigma_x$ and $\sigma_y$ , for 
several volume selections with $eB=0$ (left panels)  and for several $eB$ selections at 
infinite volume (right panels). The results are shown for $\mu_B = 0$~GeV (top panels) 
and $\mu_B=0.2$~GeV (bottom panels). A fit function is used for $eB > 0$ results.
 \label{fig:Sxy}
 }
}
\end{figure*}
The fourth term in Eq. (\ref{potential}) $ \Omega_{\bar{\psi}\psi}^{0} (T,\mu_f;\phi,\phi^{*}) $ gives the 
quark and anti-quark contributions for vanishing magnetic field. This potential can be expressed as \cite{Mao:2010},
%
\begin{eqnarray} \label{z_MF}
\Omega_{\bar{\psi}\psi}^{0} (T,\mu_f;\phi,\phi^{*})  &=& -2 T \sum_{f}  \int \dfrac{d^3 p}{(2 \pi)^3}
\\ \nn &&
\left\{ \ln \left[ 1+3\left(\phi+\phi^* e^{-\frac{(E_{f}^{0} - \mu_f)}{T}}\right)\, e^{-\frac{(E_{f}^{0} - \mu_f)}{T}}+e^{-3 \frac{(E_{f}^{0} - \mu_f)}{T}}\right] \right.
\\ \nn &&
\left.  +\ln \left[ 1+3\left(\phi^*+\phi e^{-\frac{(E_{f}^{0} + \mu_f)}{T}}\right)\, e^{-\frac{(E_{f}^{0} + \mu_f)}{T}}+e^{-3\frac{(E_{f}^{0} + \mu_f)}{T}}\right] \right\}.
\end{eqnarray} 
%
The effects of a finite volume are introduced in the PLSM via a lower momentum cut-off $p_{min}[GeV]=\pi/R[GeV]=\lambda$, 
where $R$ is the length of a cubic volume \cite{HRG_vol}.
\section{Results and discussion}
\label{sec:III}
In the following, several results are presented to illustrate the effects of finite volumes and magnetic field  strengths on 
the PLSM order parameters, thermodynamic properties and  the chiral phase transition. These results were all 
obtained with  the values for the model parameters summarized in Table~\ref{par_tab}. 
\begin{widetext}
\begin{center}
\begin{table}[hbt]
\begin{center}
 \begin{tabular}{|c|c|c|c|c|c|c|}
 \hline
  $m_{\sigma}$ (MeV) & $c$ (MeV)& $\lambda_{1} $ & $m^{2} $ ($MeV^{2}$) & $\lambda_{2} $ & $h_{x} $ ($MeV^{3}$)  & $h_{y} $ ($MeV^{3}$) \\
 \hline
  $800$ &  $4807.84$ & $13.49$ & $ -(306.26)^2$ & $46.48$ & $(120.73)^3$  & $(336.41)^3$ \\
 \hline
\end{tabular} 
\caption{Summary of the values of the PLSM parameters employed in the calculations. A detailed 
description of these parameters is given in Ref. \cite{Schaefer:pra}. 
\label{par_tab}}
\end{center}
\end{table}
\end{center}
\end{widetext}

\subsection{Order parameters}
\label{subsec:I}
%

%
%
%
\begin{figure*}[htb]
\centering{
\includegraphics[width=4.5cm,angle=-90]{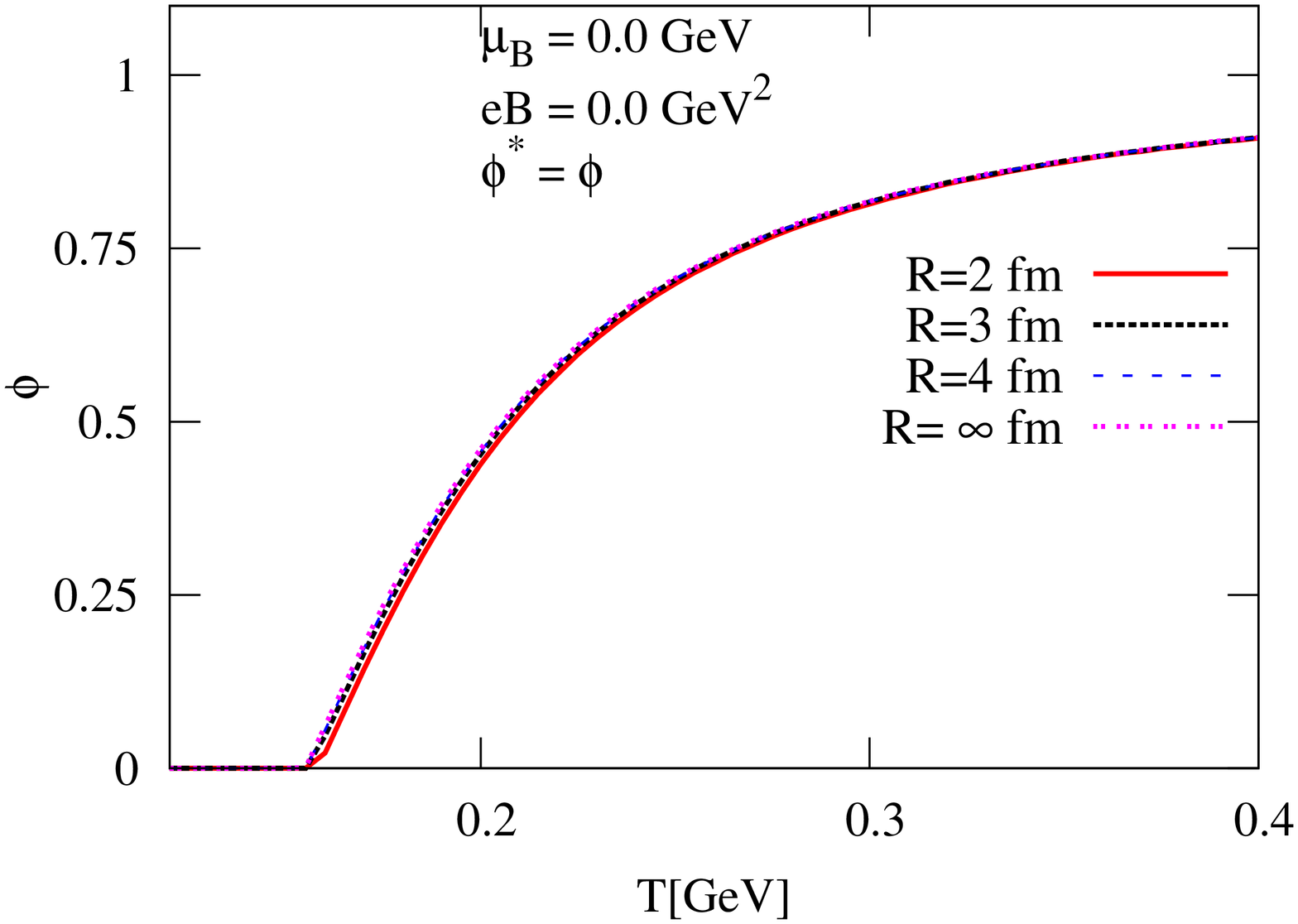}
\includegraphics[width=4.5cm,angle=-90]{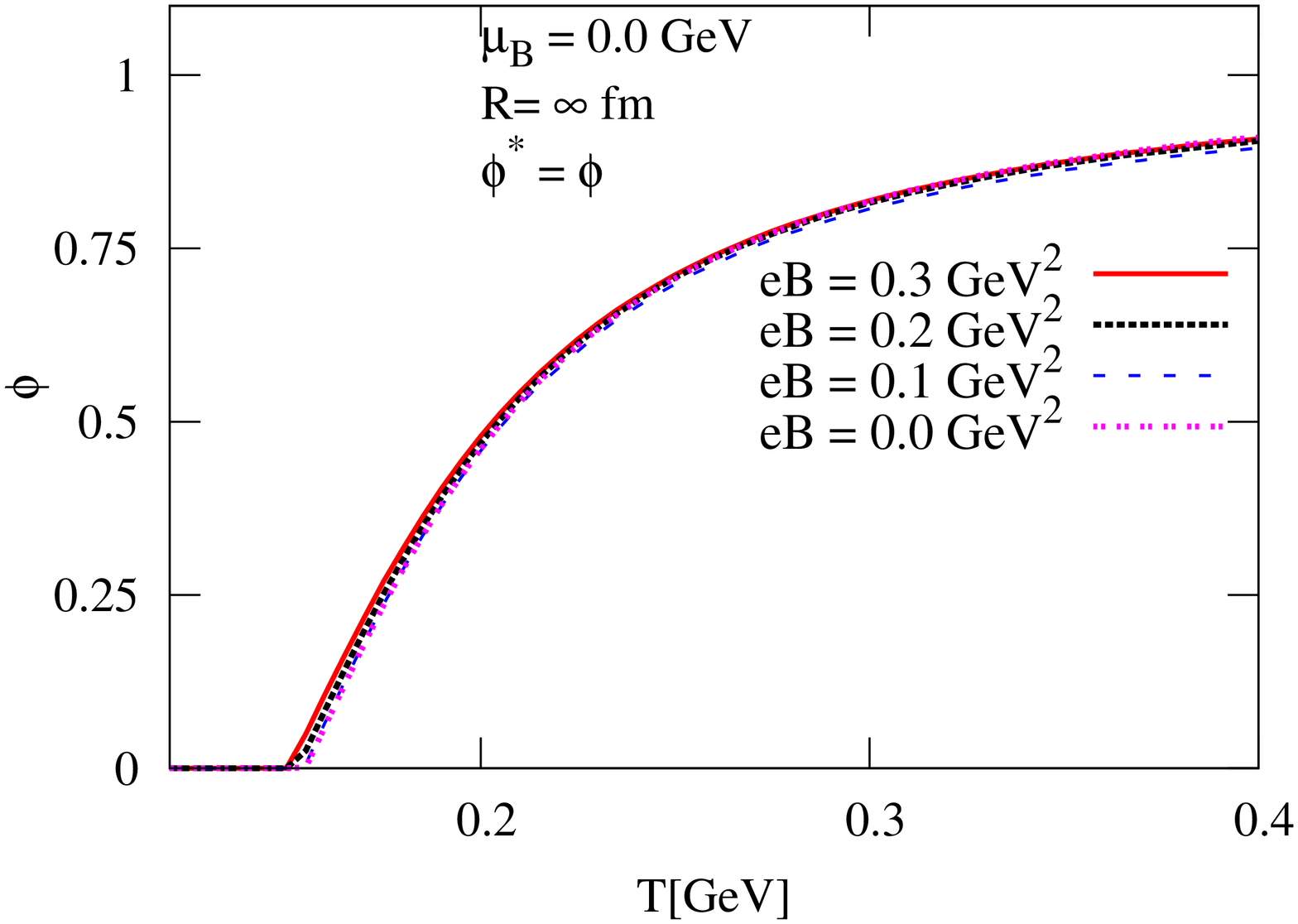}
\includegraphics[width=4.5cm,angle=-90]{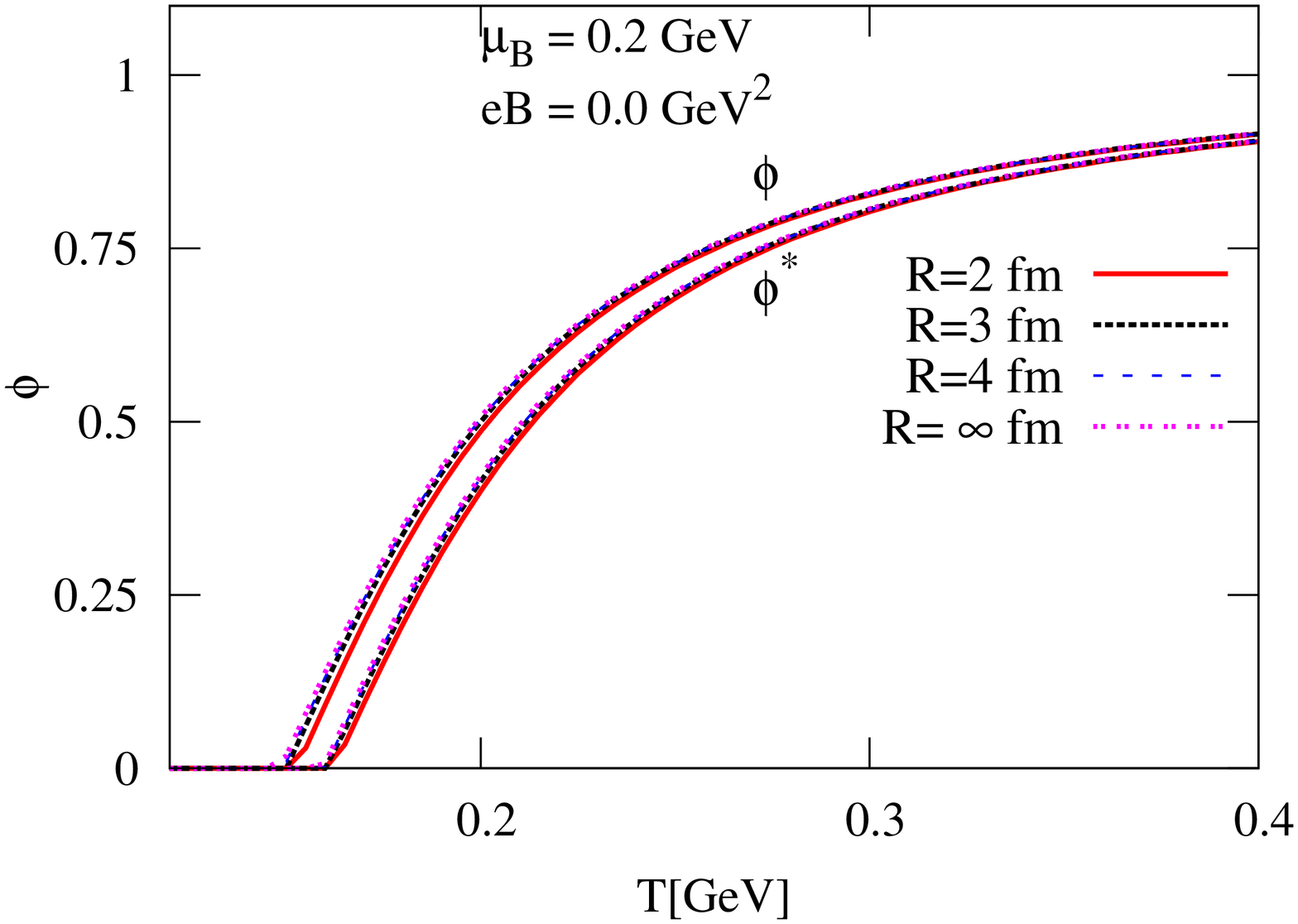}
\includegraphics[width=4.5cm,angle=-90]{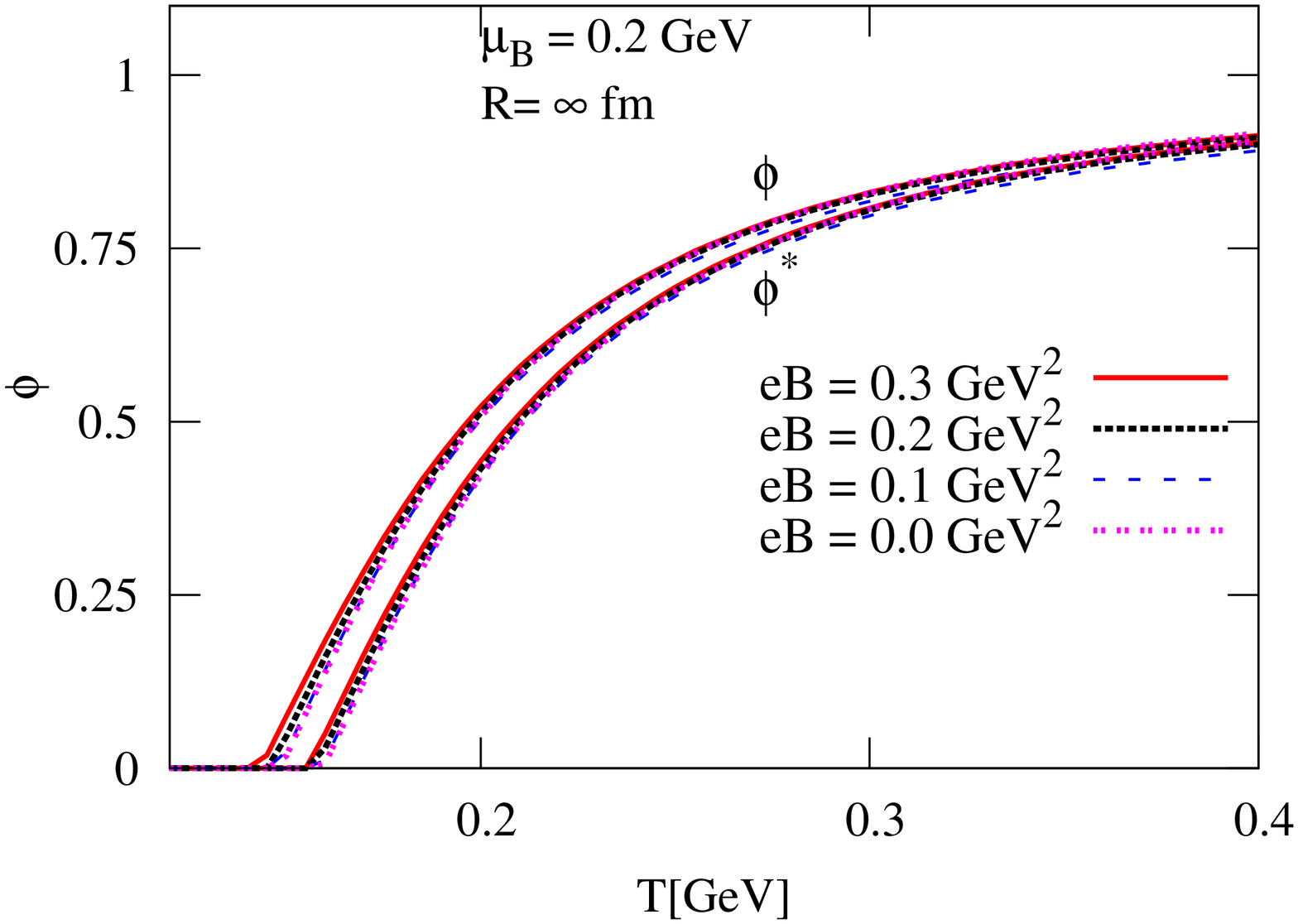}
\caption{(Color online) The same as in Fig.(\ref{fig:Sxy}) but for the two Polyakov loops.
 \label{fig:Phi}
 }
}
\end{figure*}
%
%
%
\begin{figure*}[htb]
\centering{
\includegraphics[width=4.0cm,angle=-90]{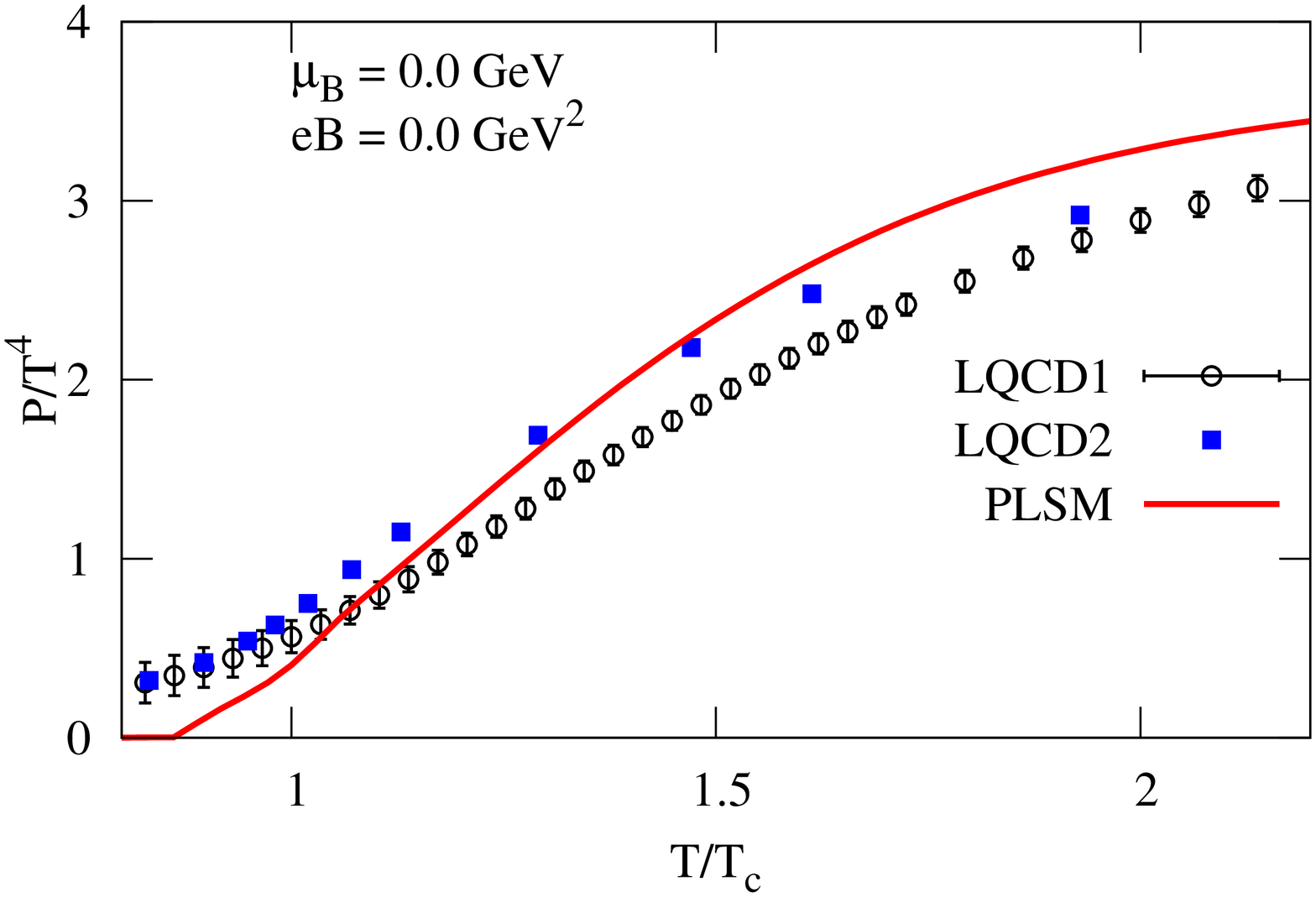}
\includegraphics[width=4.0cm,angle=-90]{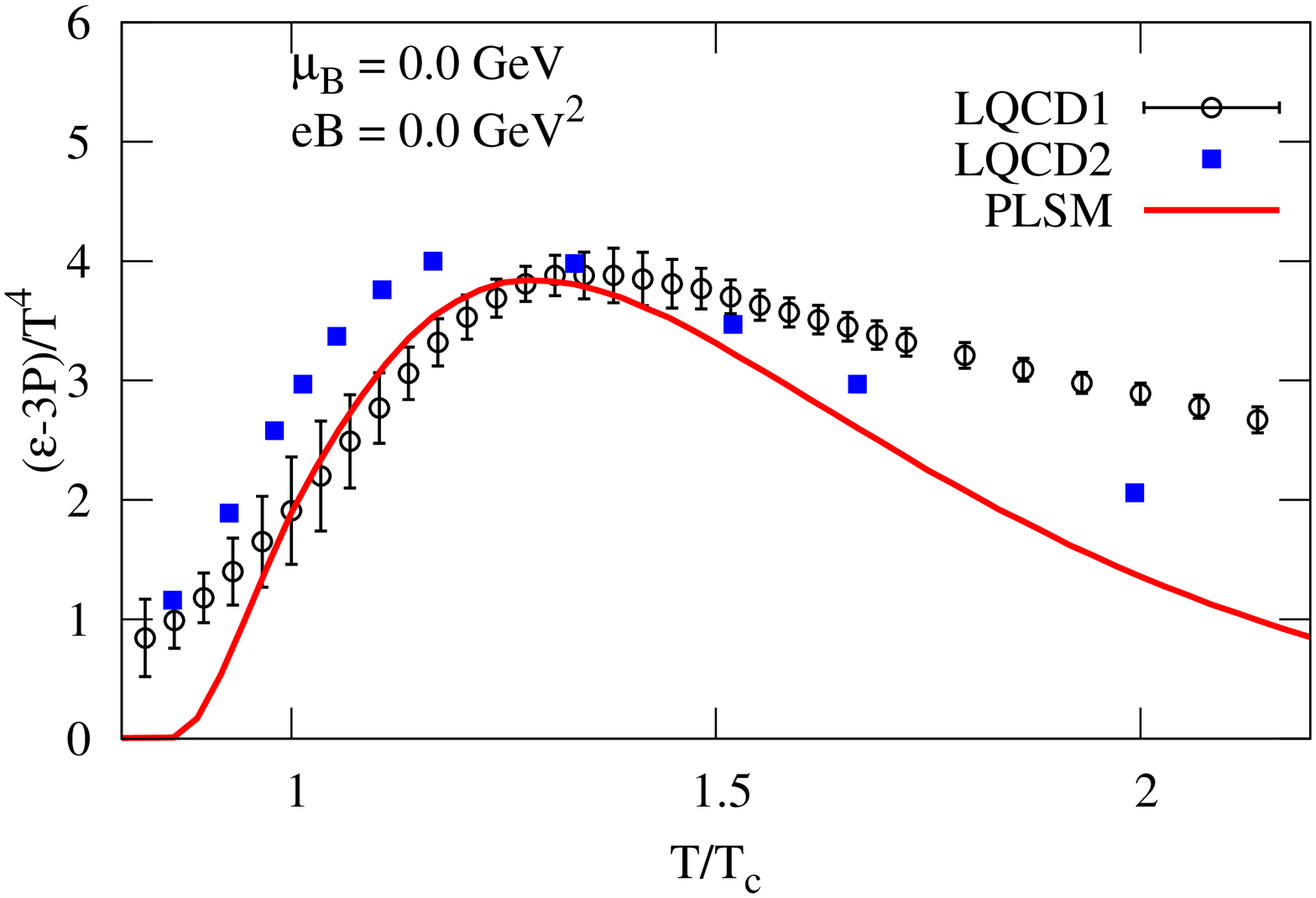}
\includegraphics[width=4.0cm,angle=-90]{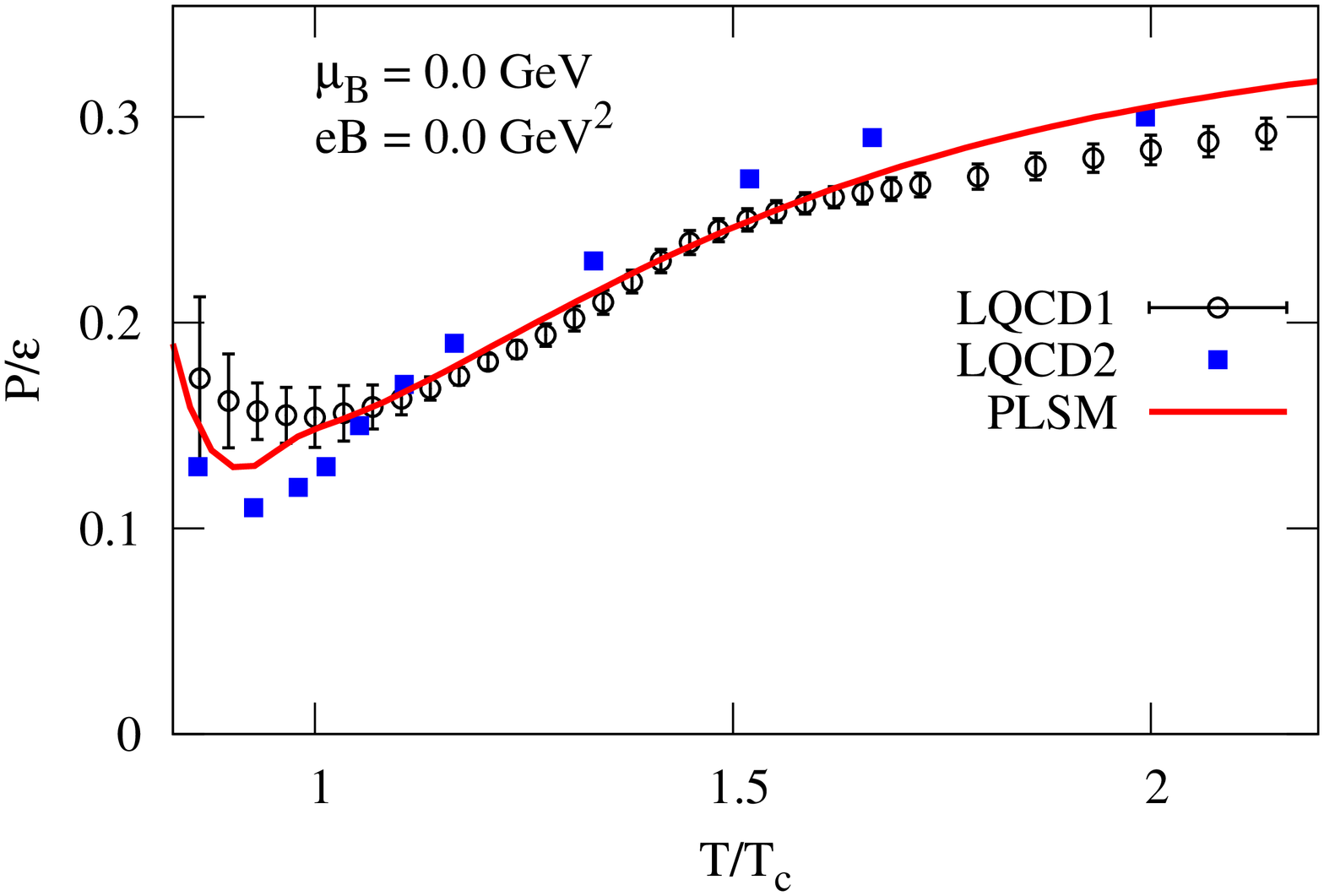}
\caption{(Color online) Comparison of  the PLSM pressure density (left panel), 
trace anomaly (middle panel) and $P/\epsilon$ (right panel) to results from 
LQCD. The comparisons are made for $\mu_B = 0$; the solid lines indicate 
PLSM results and the open and closed symbols indicate LQCD results from 
Refs.~ \cite{QCDL} and \cite{QCD10} respectively.
 \label{fig:LQCD}
 }
}
\end{figure*}

Figure~\ref{fig:Sxy} shows the temperature dependence of the two chiral condensates ($\sigma_y$ and $\sigma_x$) 
for different volume and magnetic field selections for two values of $\mu_B$. The left panels show that both chiral 
condensates increase as the system volume is decreased, albeit with much larger sensitivity for the non-strange chiral 
condensates ($\sigma_x$). 
The right panels indicate an opposite trend for increasing magnetic field strength,
again with with a larger sensitivity for $\sigma_x$.
Fig. \ref{fig:Phi} shows the corresponding temperature dependence of the two Polyakov loops ($\phi$ and $\phi^*$) 
for the same volume, magnetic field and $\mu_B$ selections. For $\mu_B = 0$~GeV we observe that $\phi$ = $\phi^*$  and 
both order parametes show very little, if any, dependence on the volume and the magnetic field strength.
For $\mu_B = 0.2$~GeV, a weak dependence, with trends similar to those in the bottom panels of Fig.~\ref{fig:Sxy}, can be 
observed.

%
%
%
\begin{figure*}[htb]
\centering{
\includegraphics[width=5.cm,angle=-90]{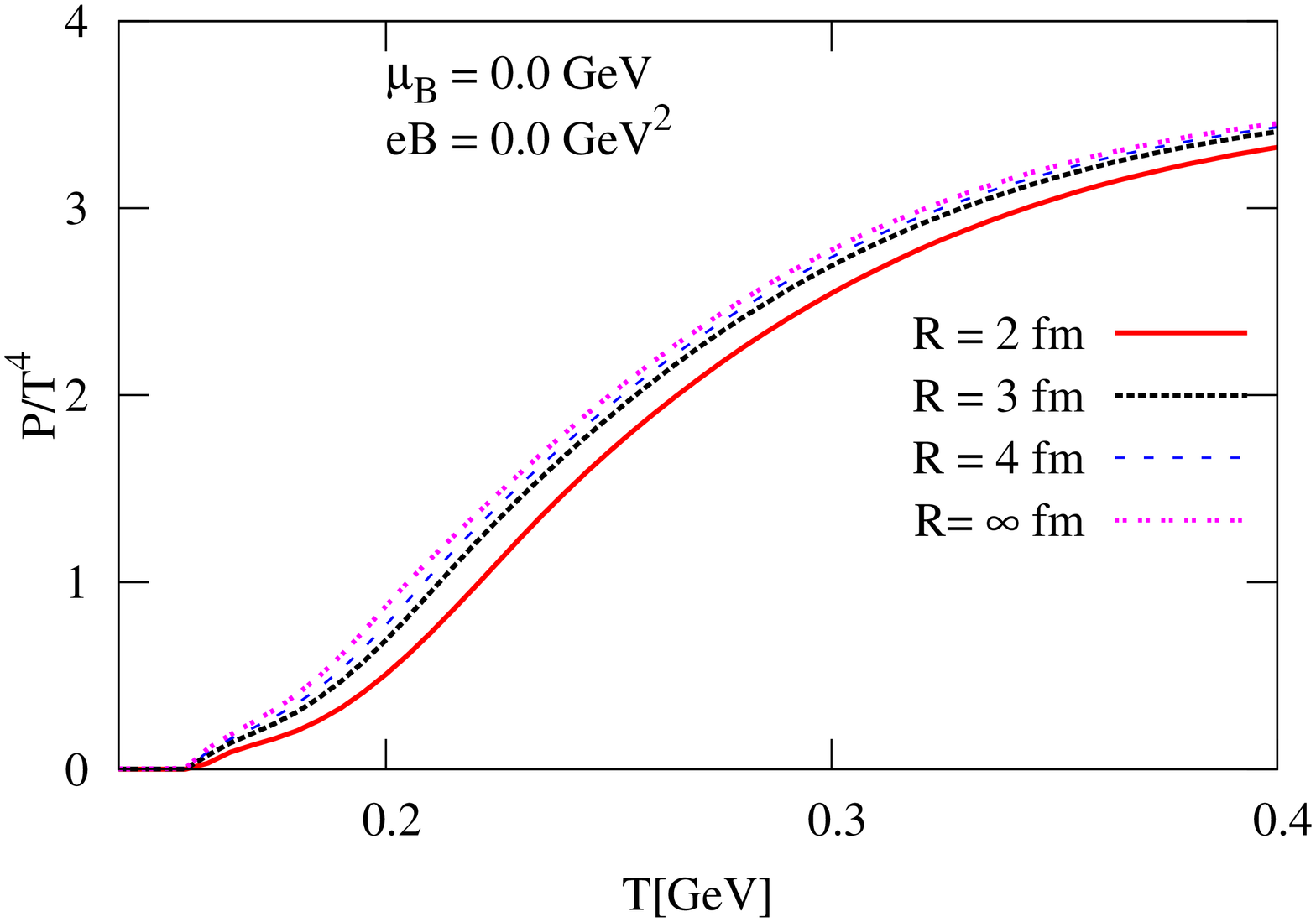}
\includegraphics[width=5.cm,angle=-90]{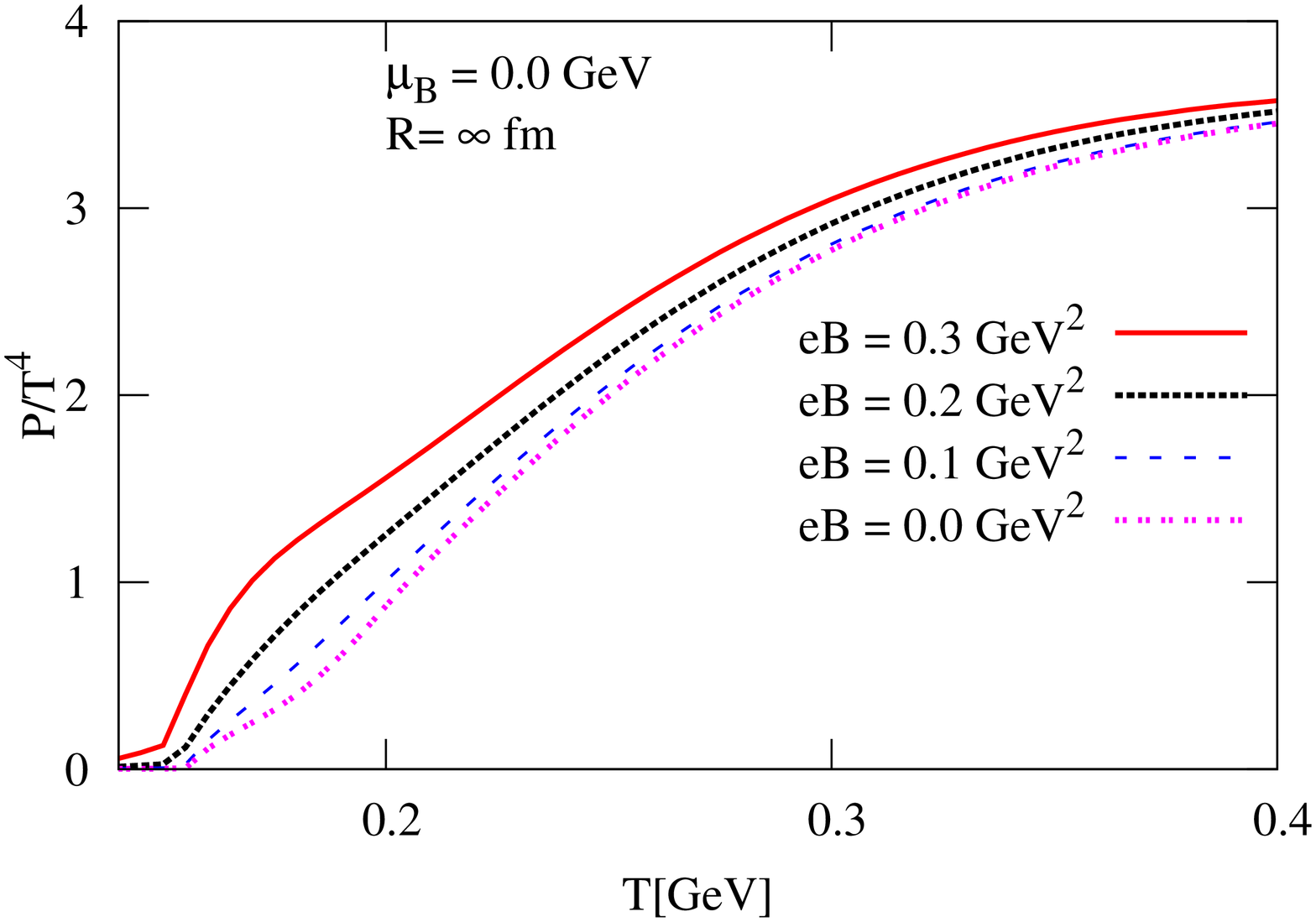}
\includegraphics[width=5.cm,angle=-90]{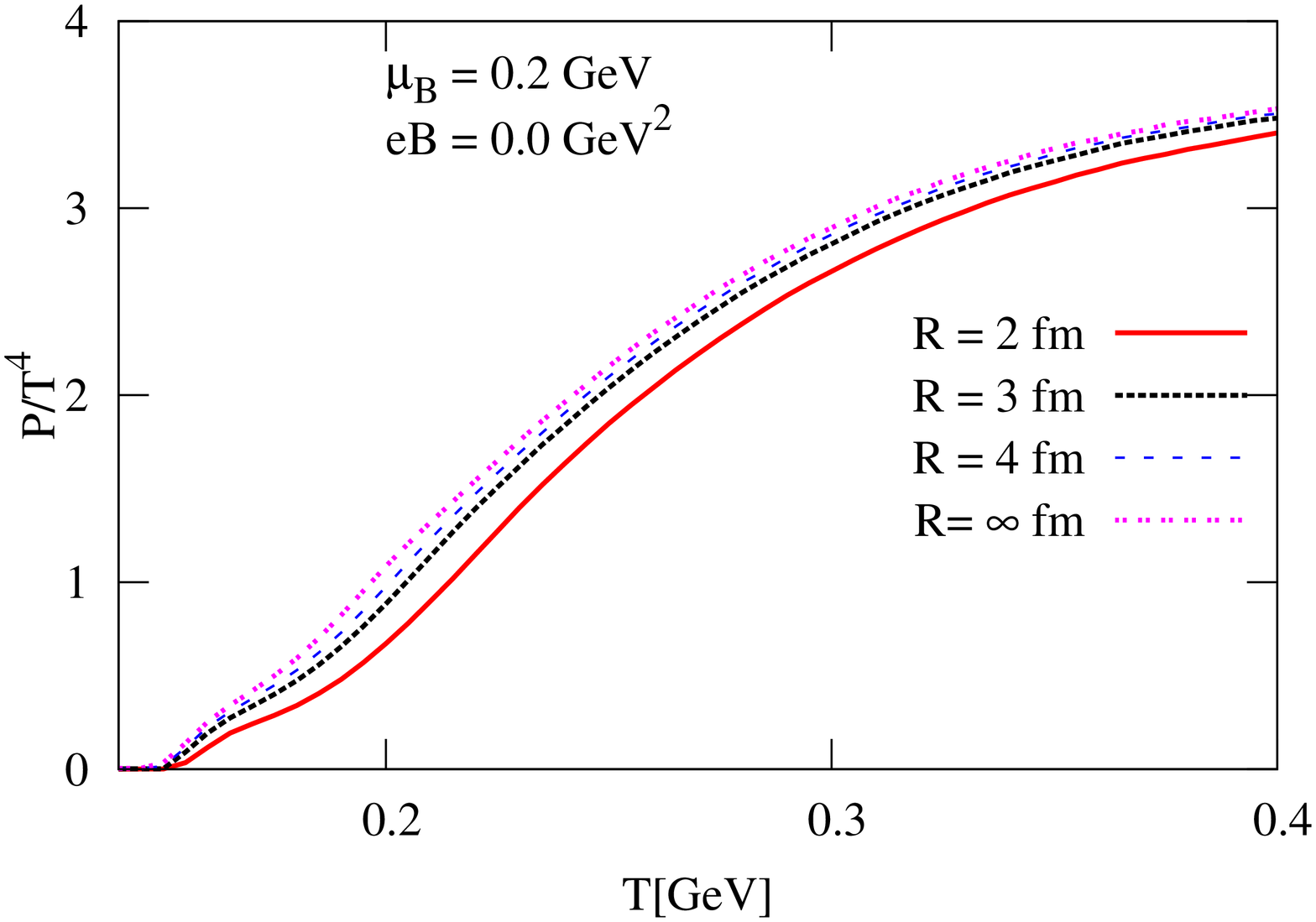}
\includegraphics[width=5.cm,angle=-90]{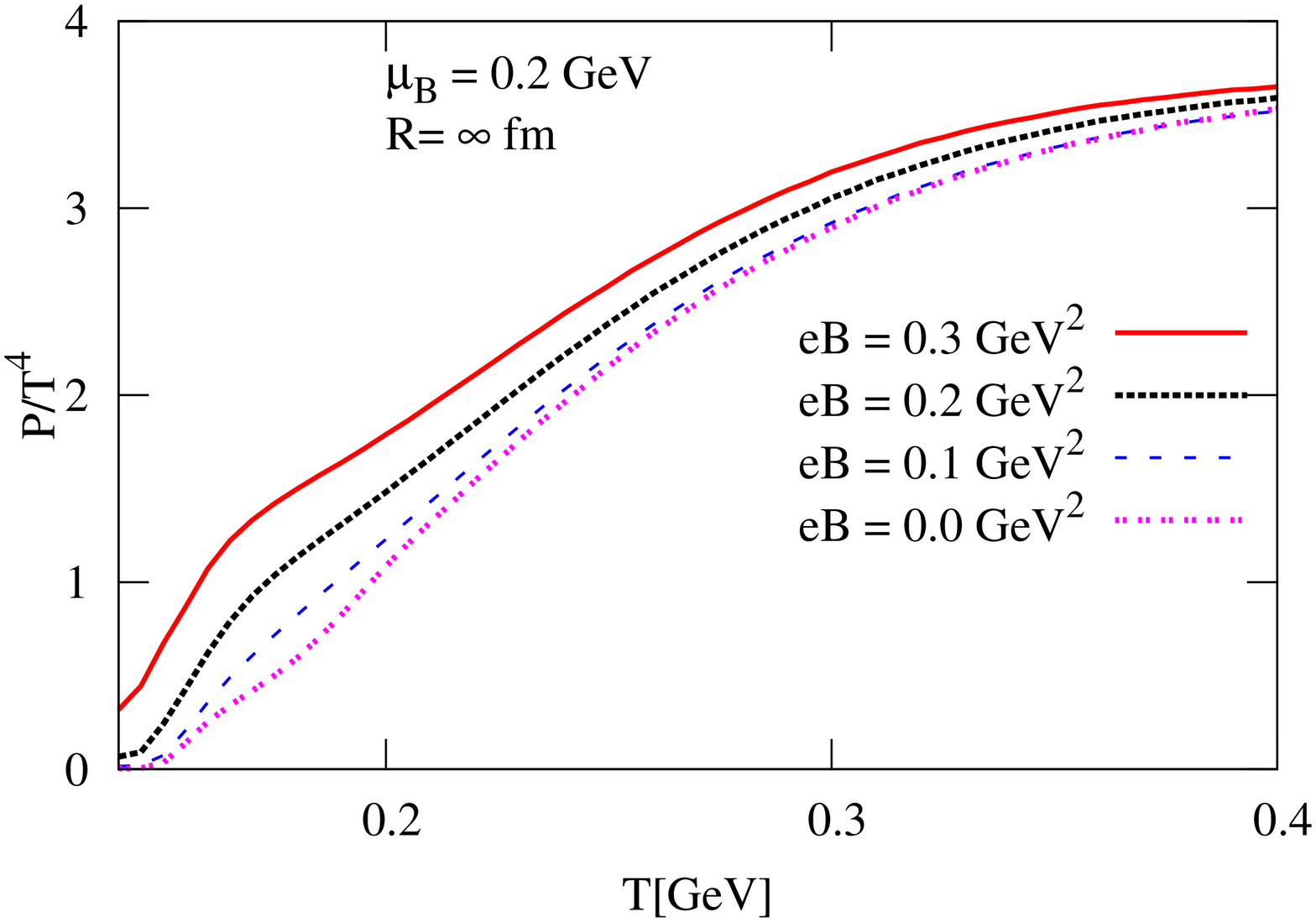}
\caption{(Color online) Temperature dependence of the normalized pressure , for 
several volume selections with $eB=0$ (left panels)  and for several $eB$ selections at 
infinite volume (right panels). Results are shown for $\mu_B = 0$~GeV (top panels) 
and $\mu_B=0.2$~GeV (bottom panels).
 \label{fig:Pr}.
 }
}
\end{figure*}

\subsection{Thermodynamics properties}
\label{subsec:II}
%
%
%
\begin{figure*}[htb]
\centering{
\includegraphics[width=5.cm,angle=-90]{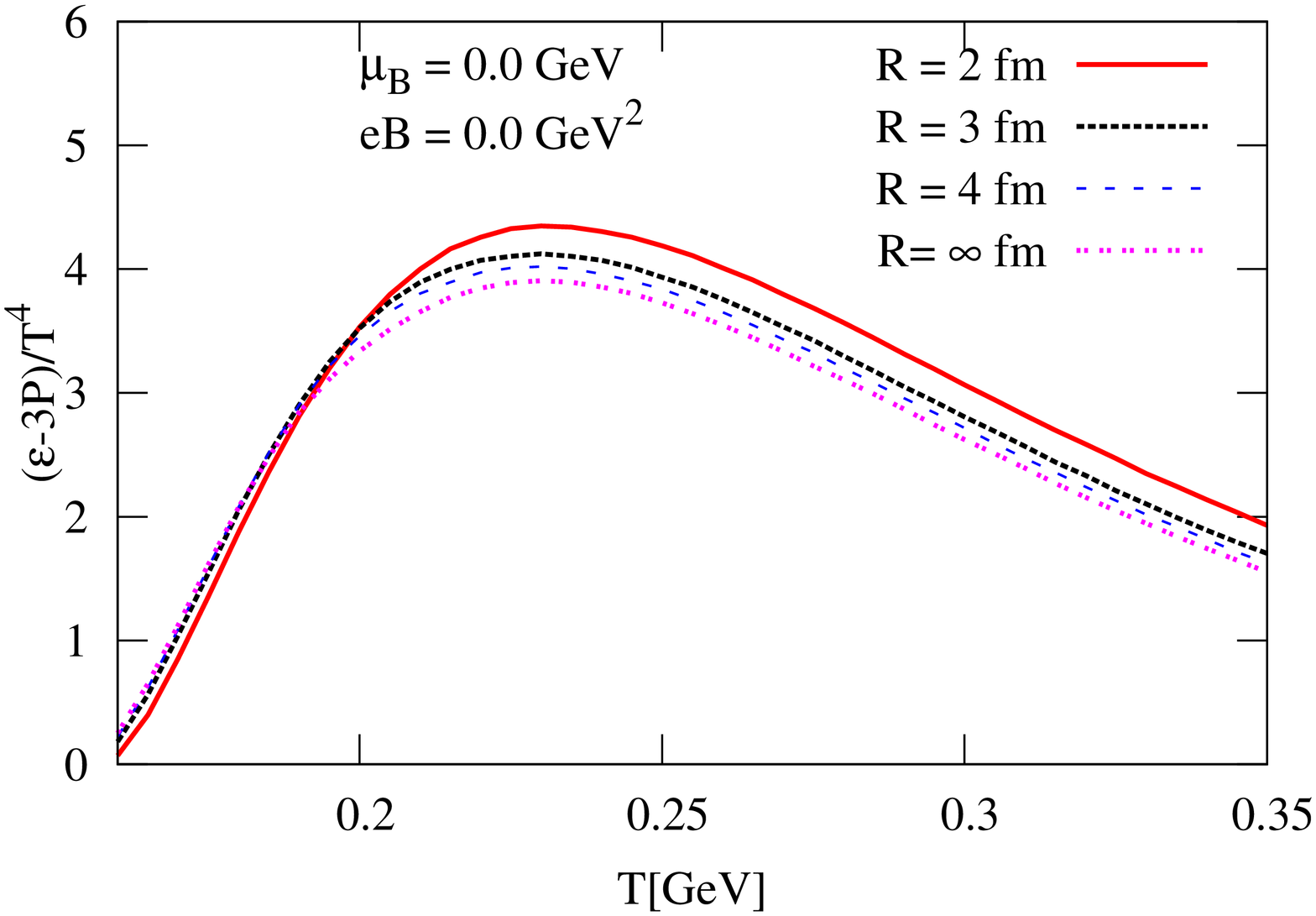}
\includegraphics[width=5.cm,angle=-90]{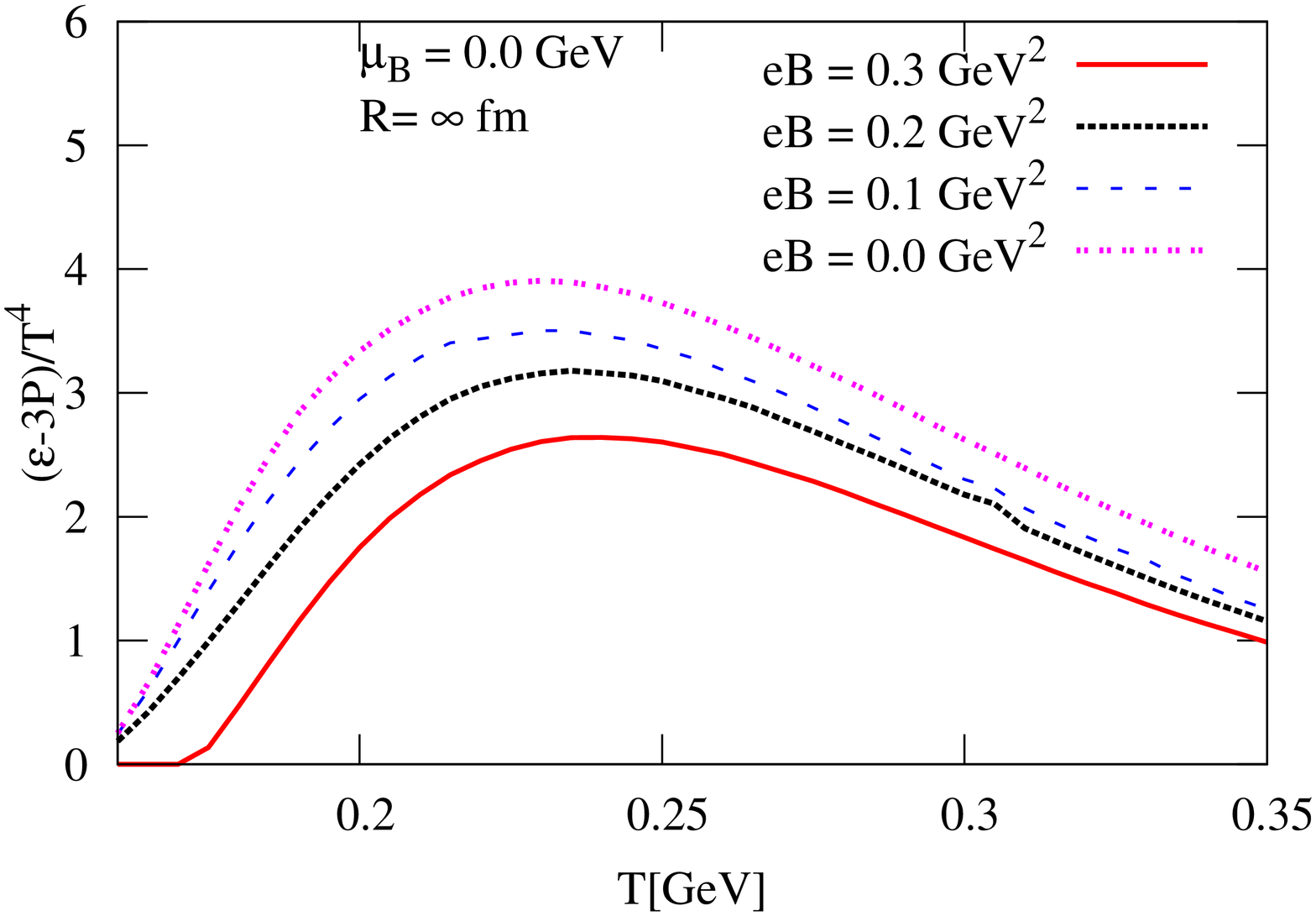}
\includegraphics[width=5.cm,angle=-90]{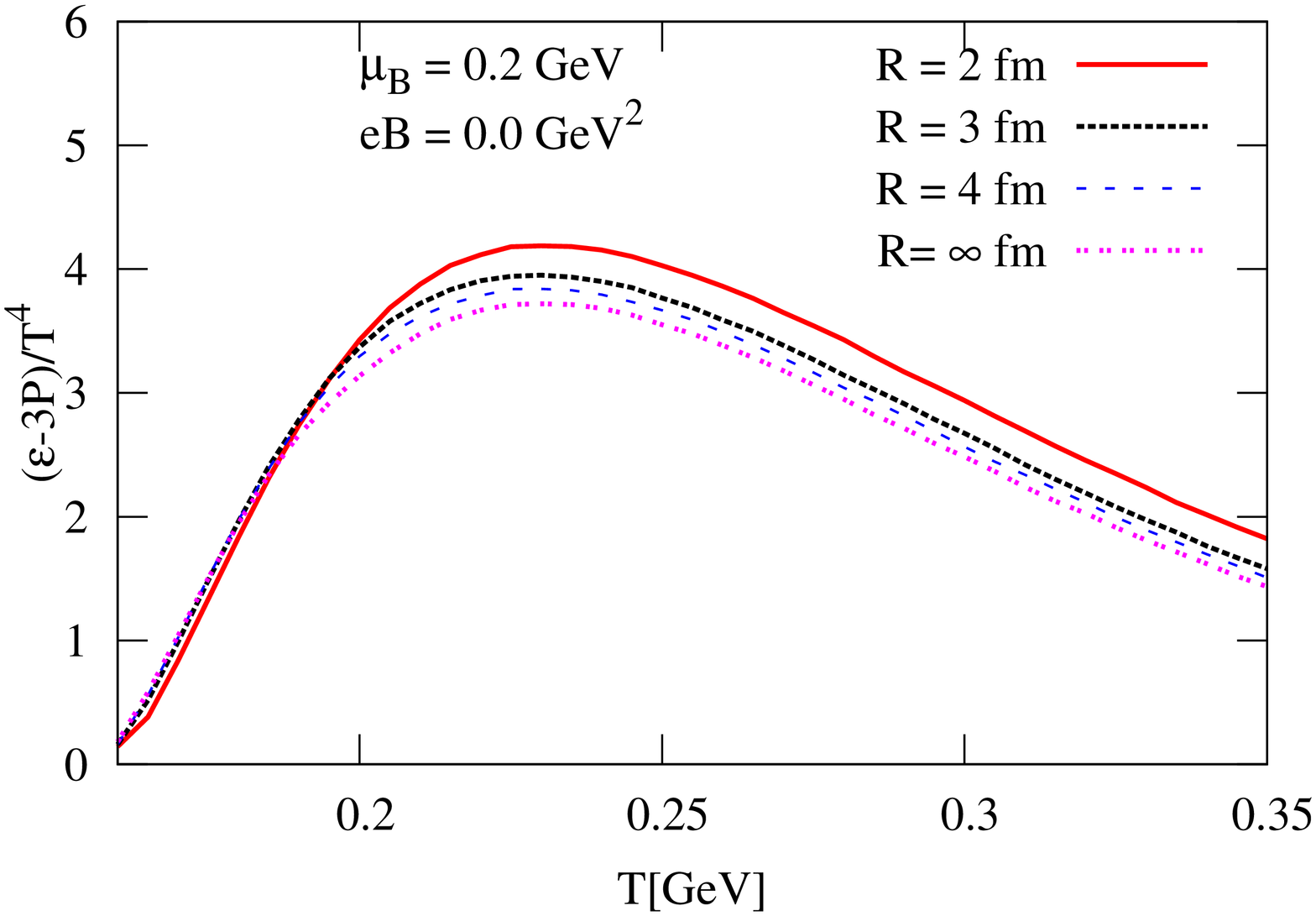}
\includegraphics[width=5.cm,angle=-90]{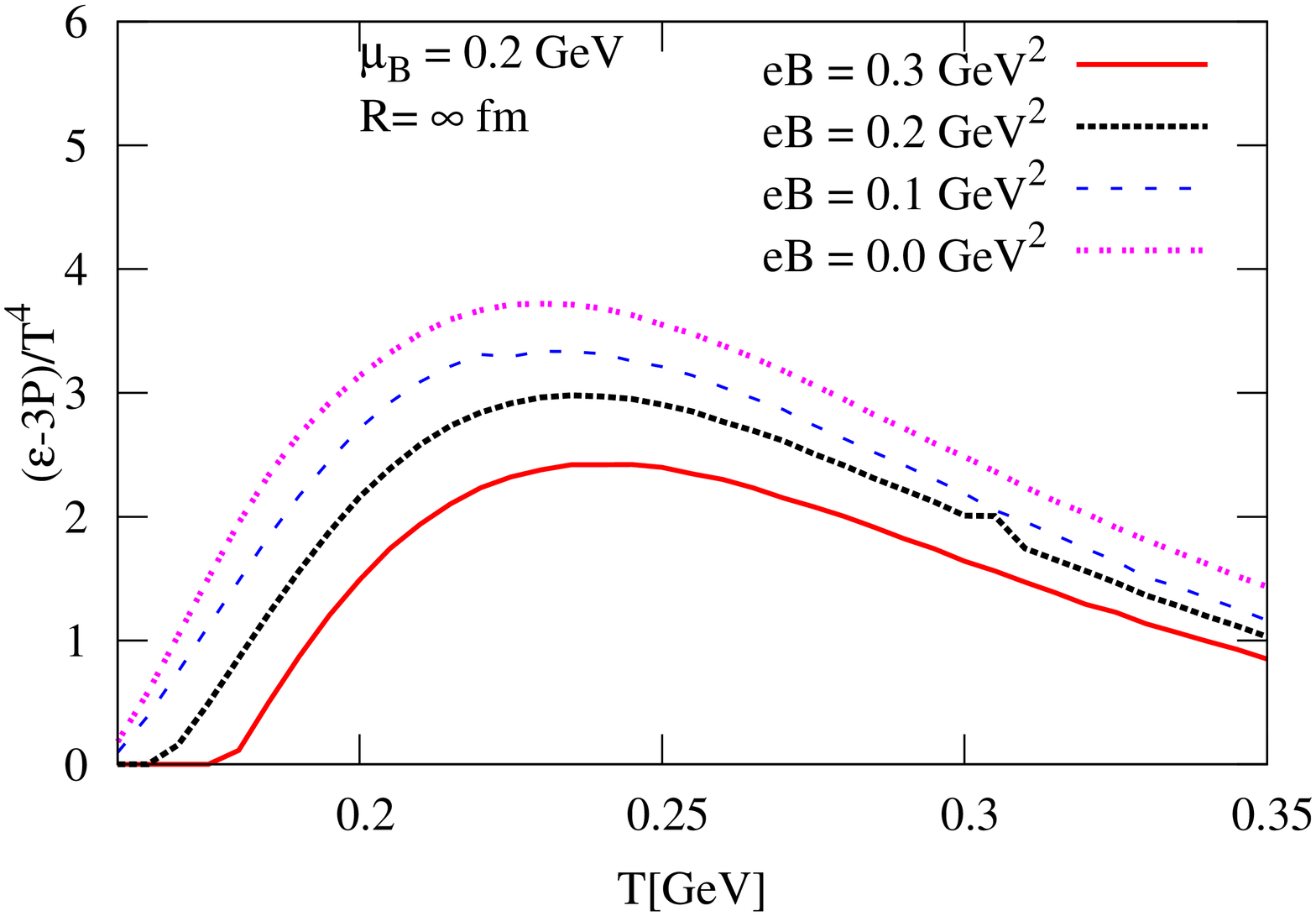}
\caption{(Color online)  Temperature dependence of the normalized trace anomaly for 
several volume selections with $eB= 0$~GeV$^2$ (left panels) and for several $eB$ selections at 
infinite volume (right panels). Results are shown for $\mu_B = 0$~GeV (top panels) 
and $\mu_B=0.2$~GeV (bottom panels).
 \label{fig:Trec}
 }
}
\end{figure*}
%
%
\begin{figure*}[htb]
\centering{
\includegraphics[width=5.2cm,angle=-90]{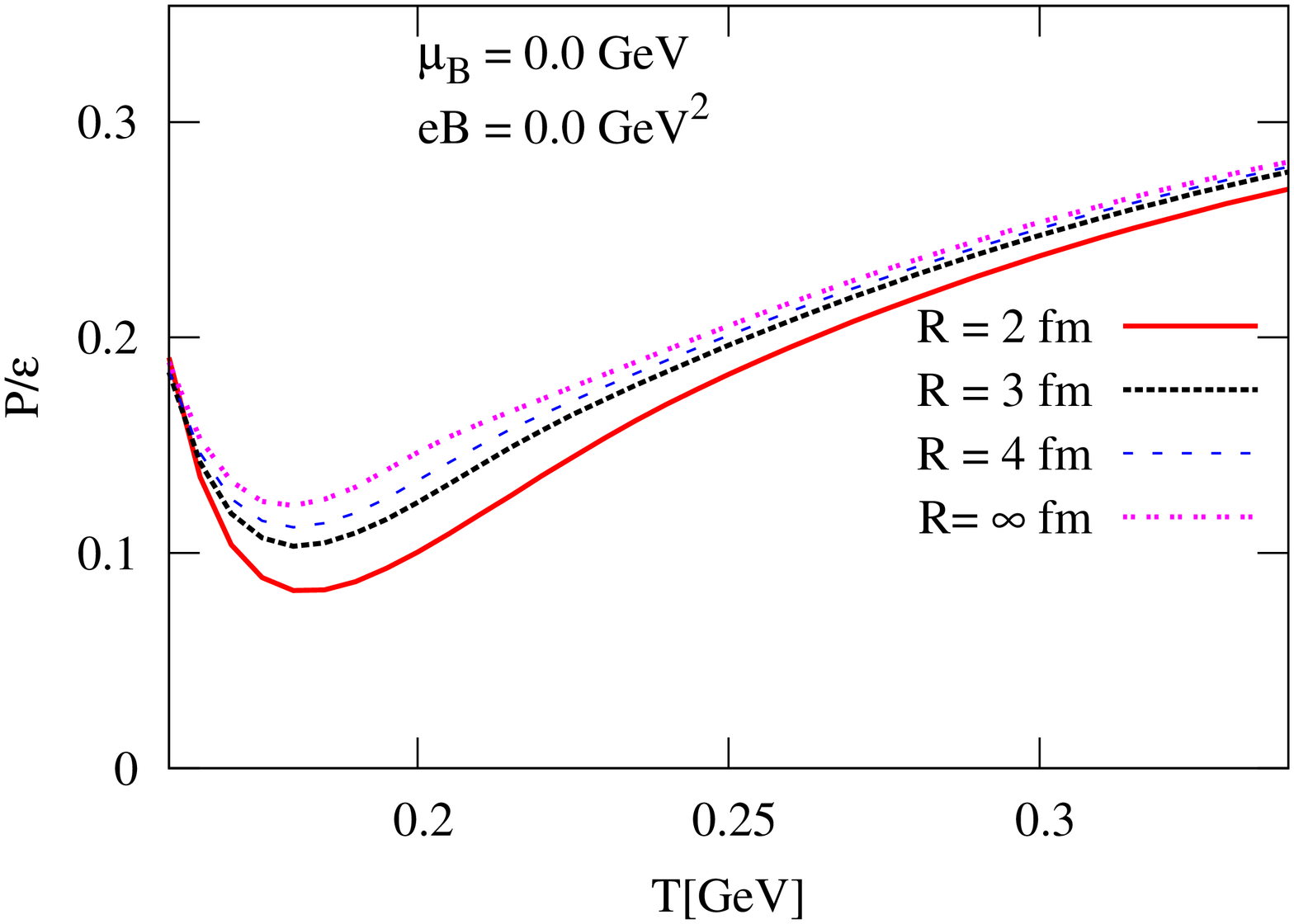}
\includegraphics[width=5.cm,angle=-90]{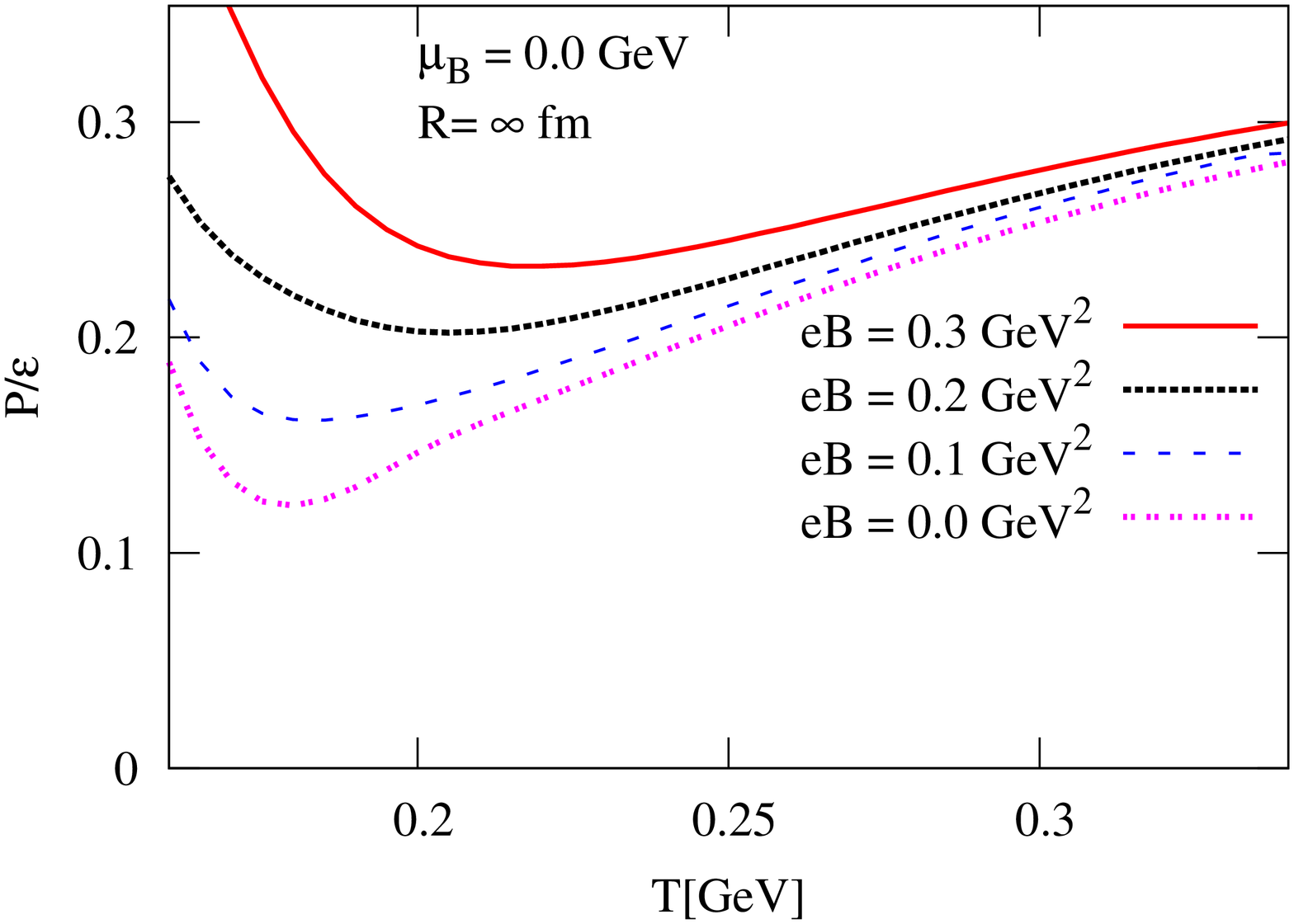}
\includegraphics[width=5.cm,angle=-90]{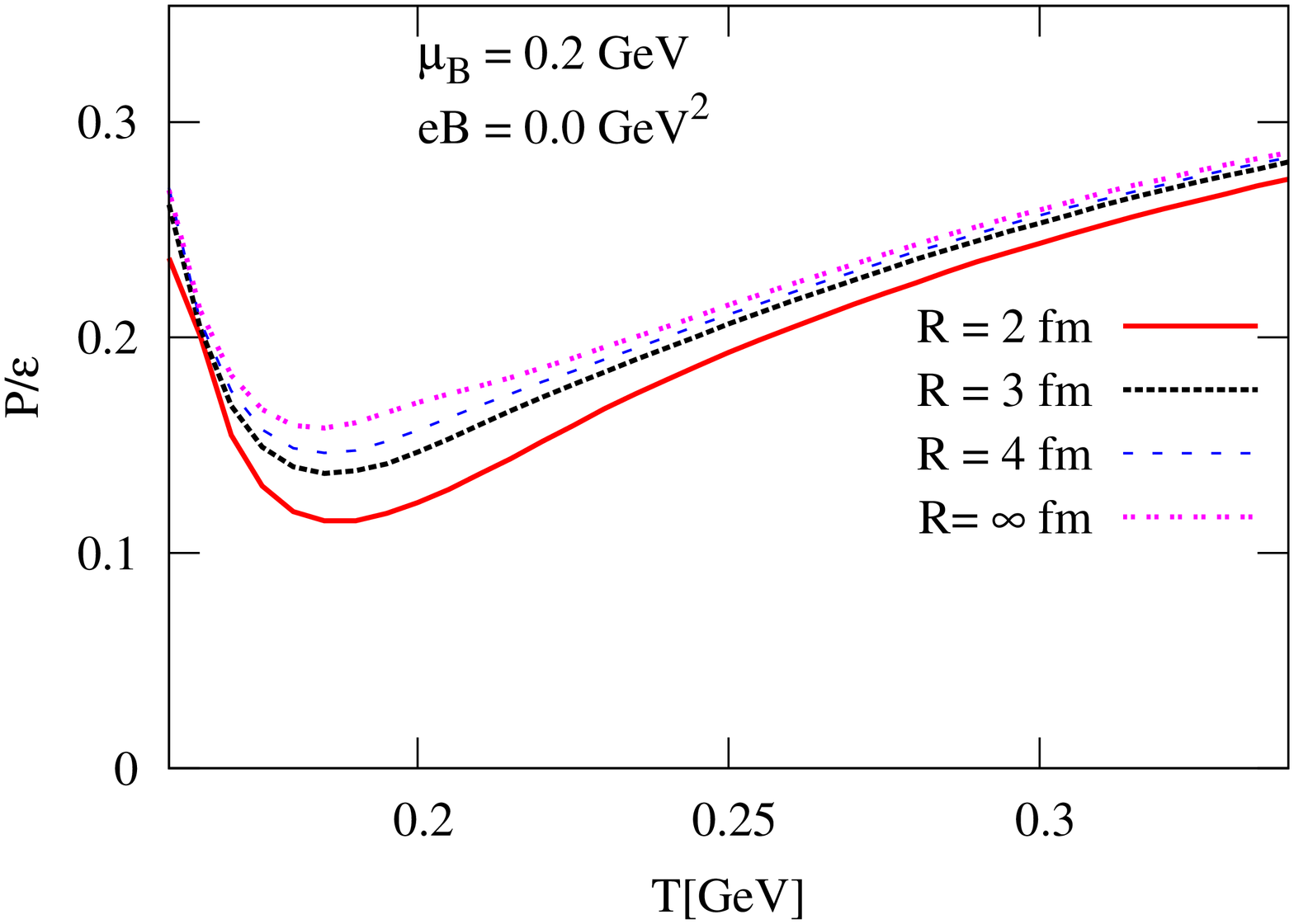}
\includegraphics[width=5.cm,angle=-90]{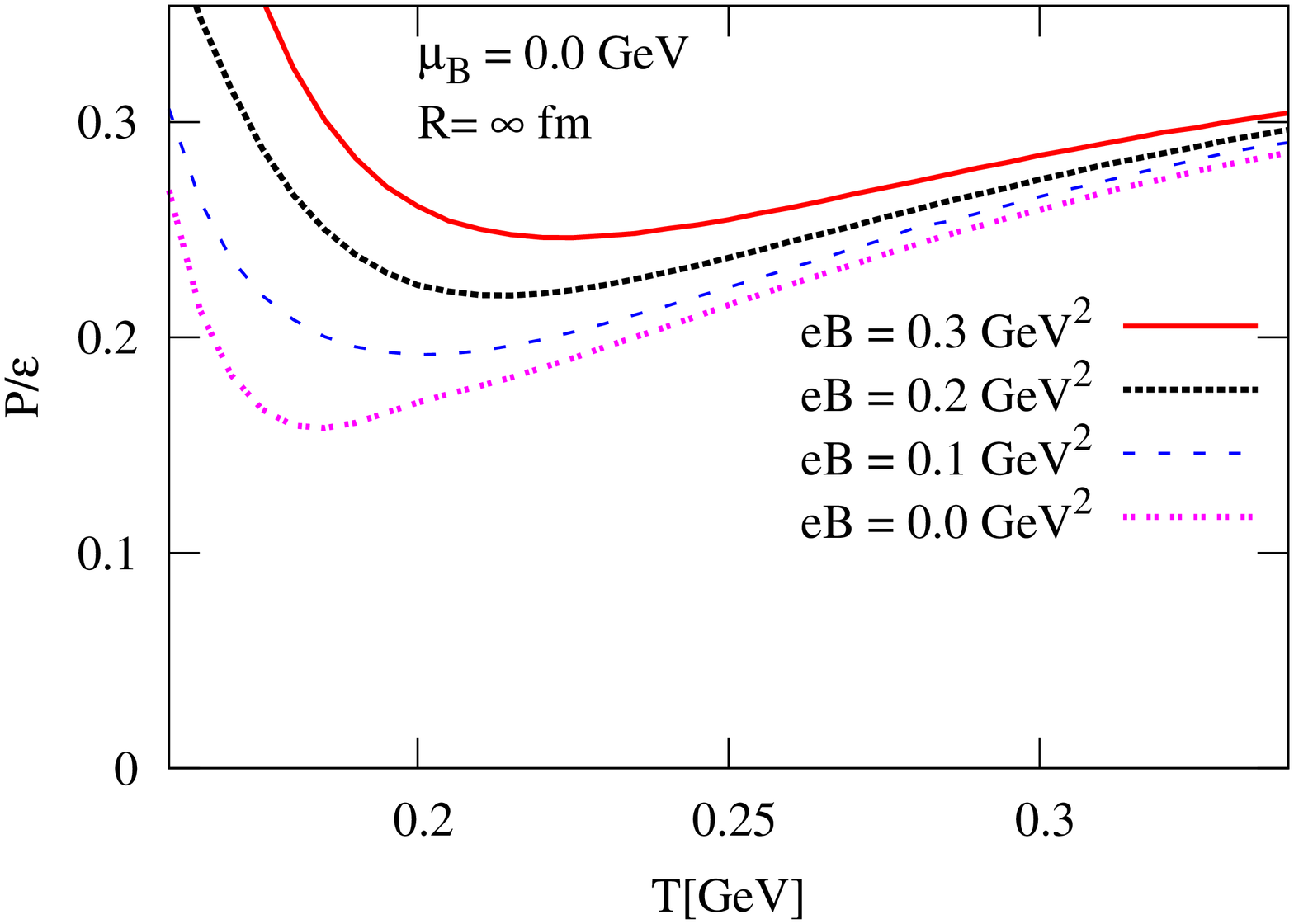}
\caption{(Color online)  $P/\epsilon$ vs. $T$ for several volume selections 
with $eB= 0$~GeV$^2$ (left panels) and for several $eB$ selections at 
infinite volume (right panels). Results are shown for $\mu_B = 0$~GeV (top panels) 
and $\mu_B=0.2$~GeV (bottom panels).
 \label{fig:Cs2}
 }
}
\end{figure*}
%
%
%
\begin{figure*}[htb]
\centering{
\includegraphics[width=5.cm,angle=-90]{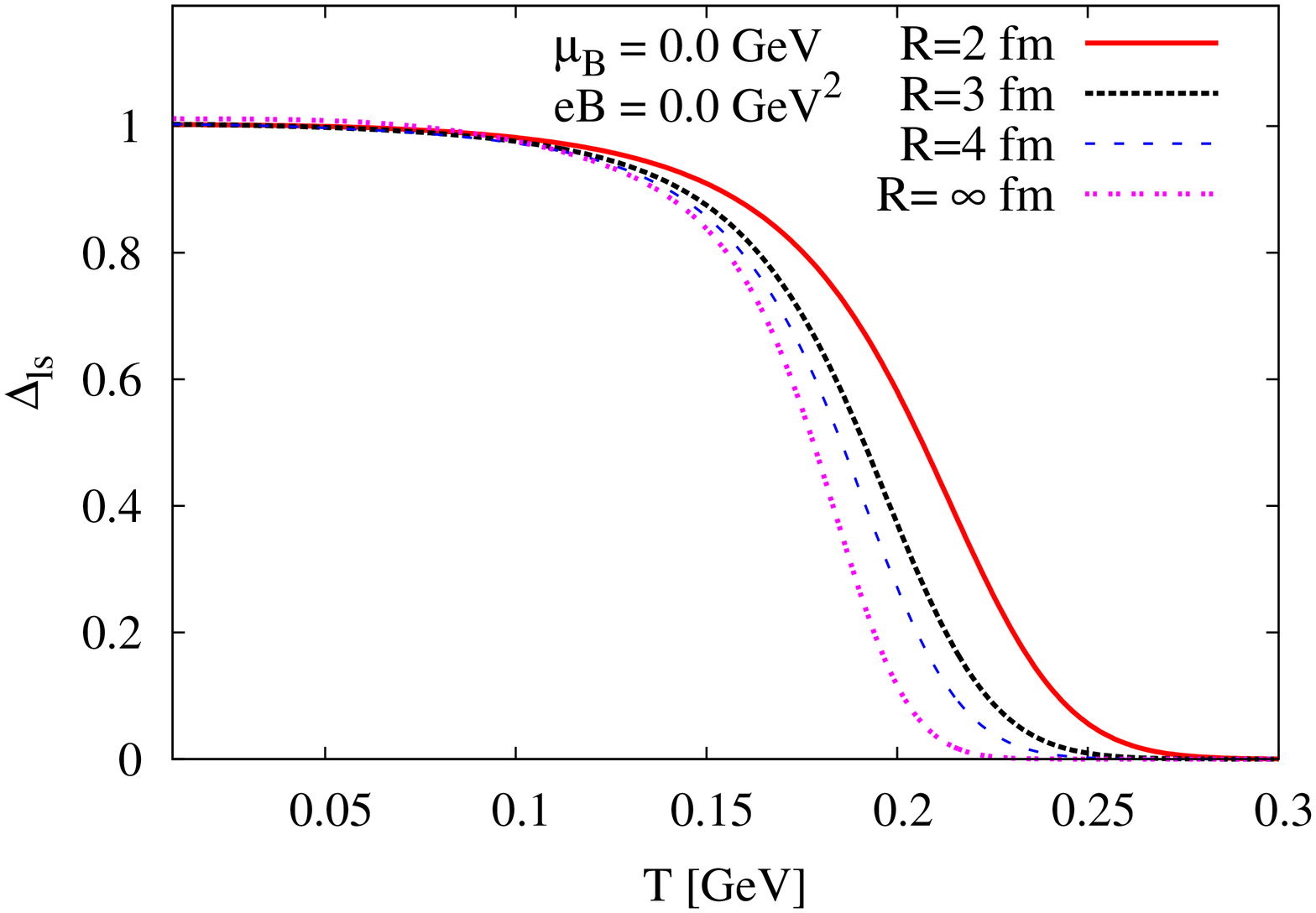}
\includegraphics[width=5.cm,angle=-90]{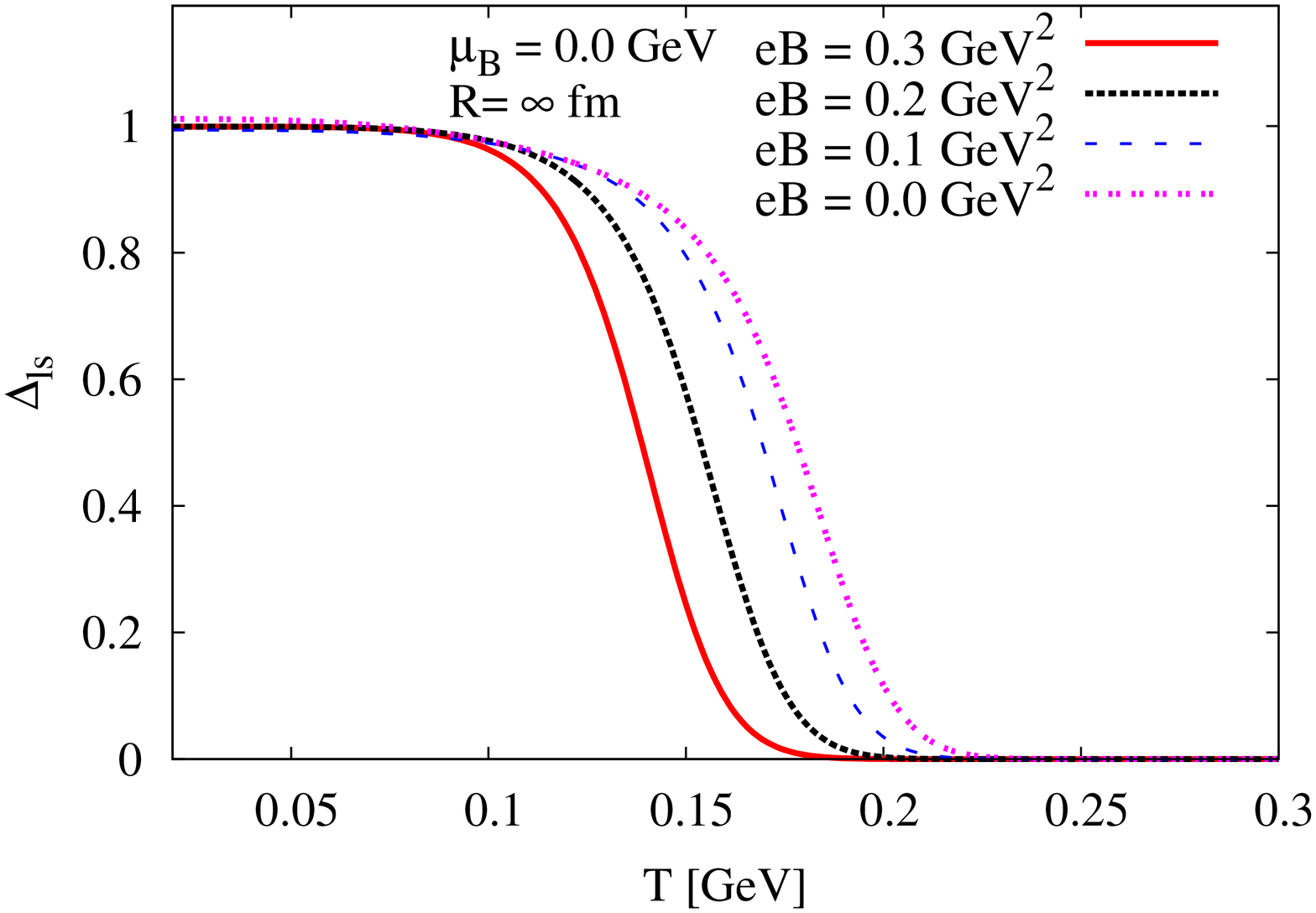}
\includegraphics[width=5.cm,angle=-90]{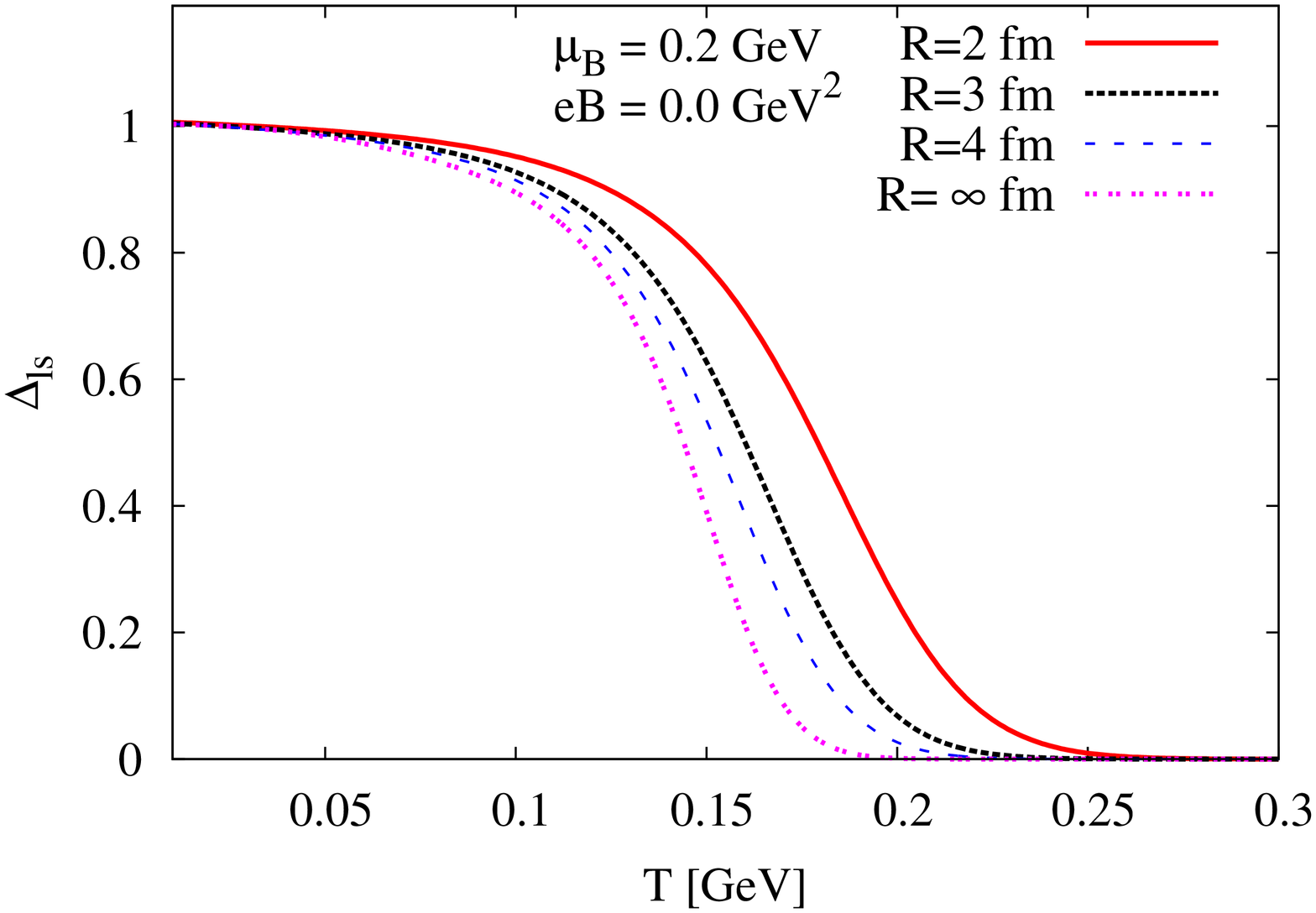}
\includegraphics[width=5.cm,angle=-90]{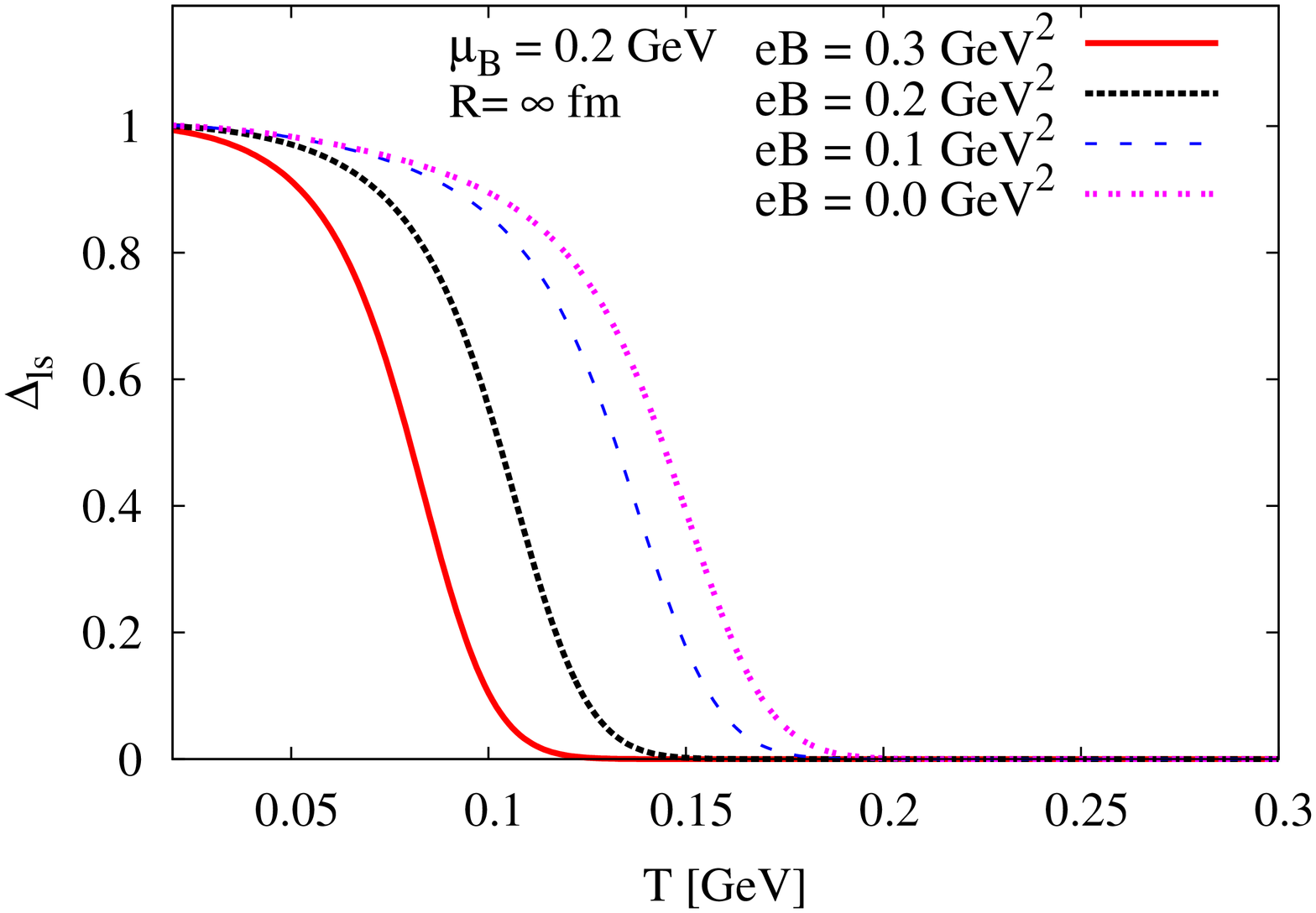}
\caption{(Color online) Temperature dependence of the net-difference condensates $\Delta_{ls}$, for 
several volume selections with $eB=0$ (left panels)  and for several $eB$ selections at 
infinite volume (right panels). The results are shown for $\mu_B = 0$~GeV (top panels) 
and $\mu_B=0.2$~GeV (bottom panels).
 \label{fig:Delta}
 }
}
\end{figure*}
%
%
%
\begin{figure*}[htb]
\centering{
\includegraphics[width=5.cm,angle=-90]{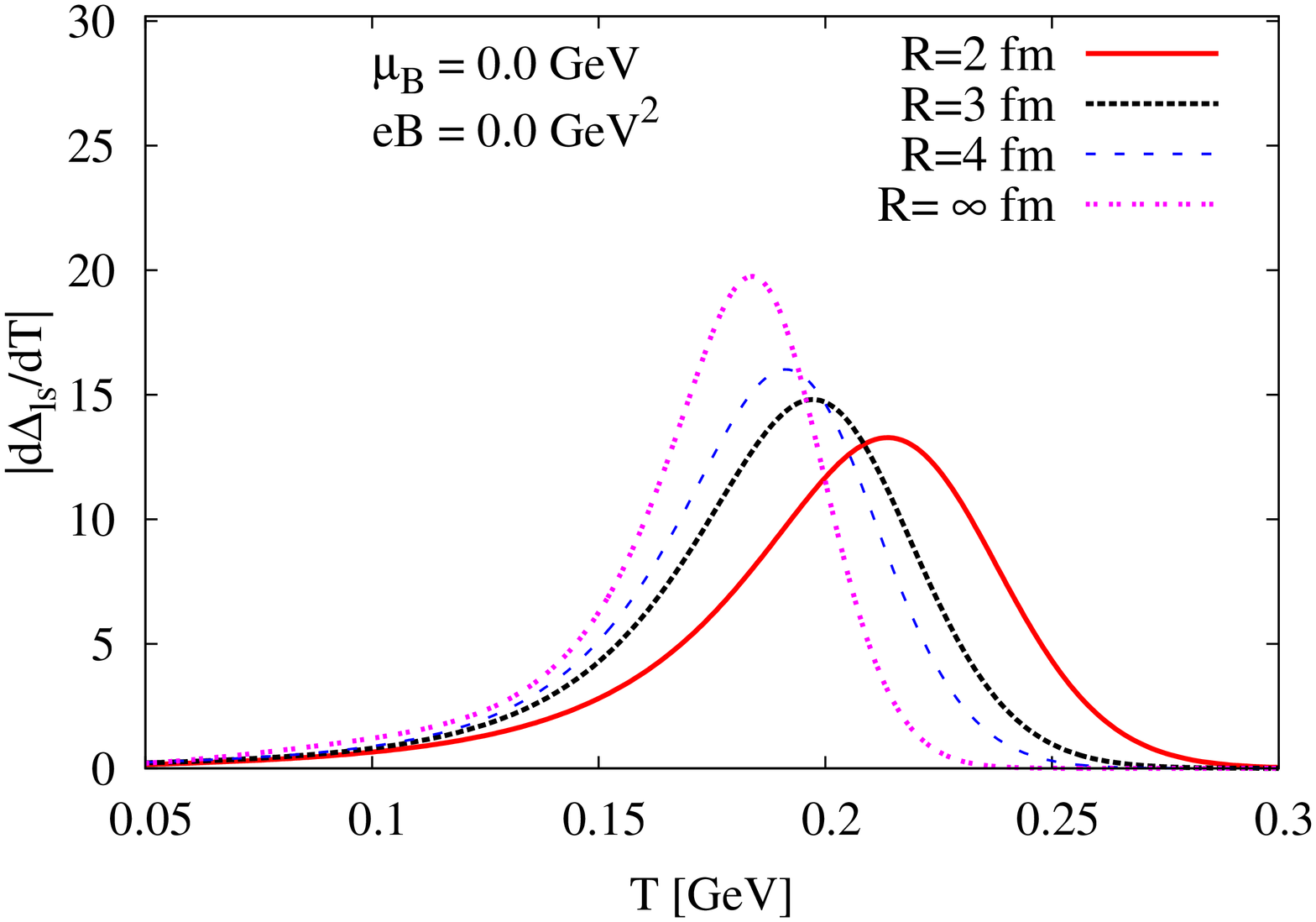}
\includegraphics[width=5.cm,angle=-90]{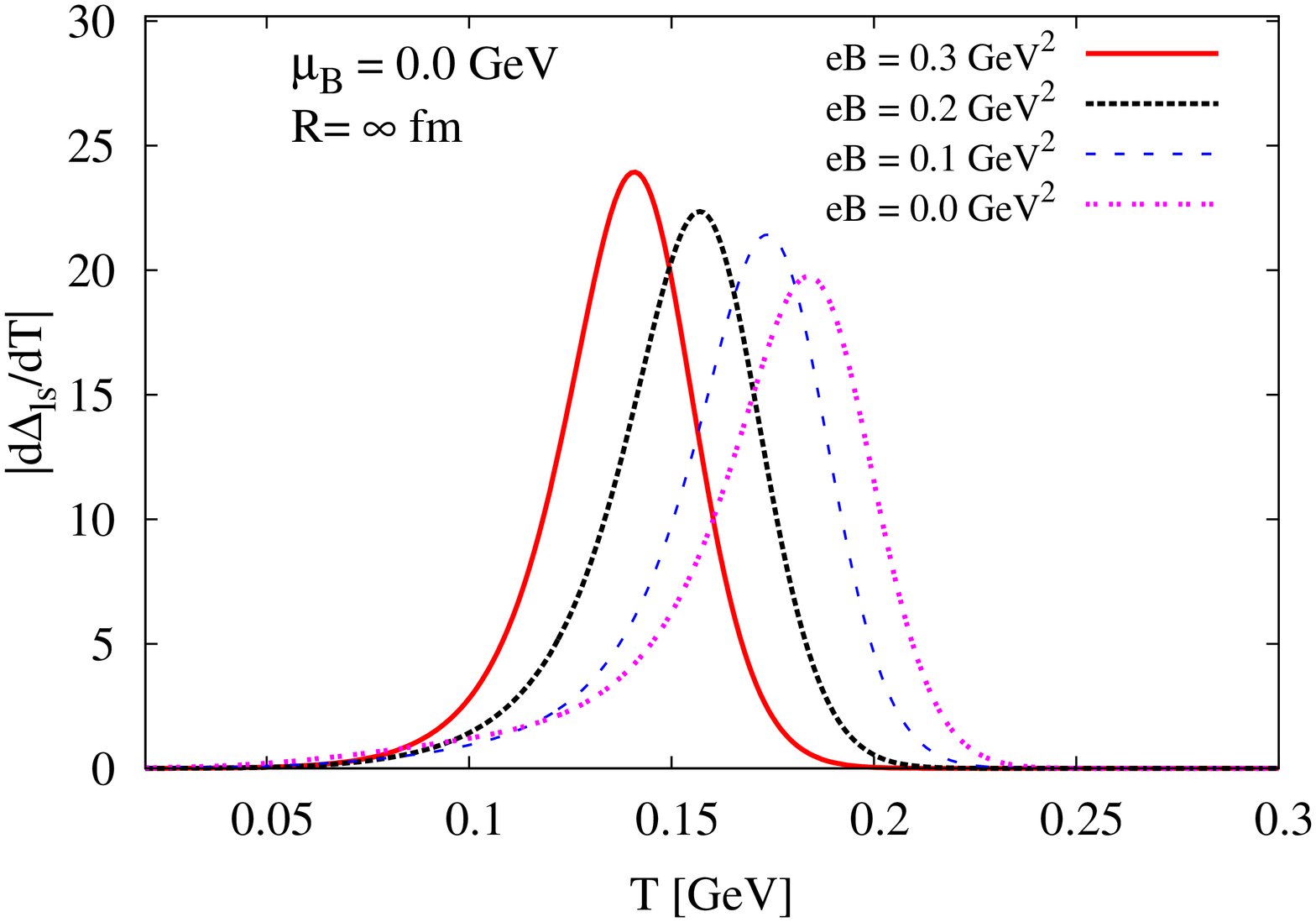}
\includegraphics[width=5.cm,angle=-90]{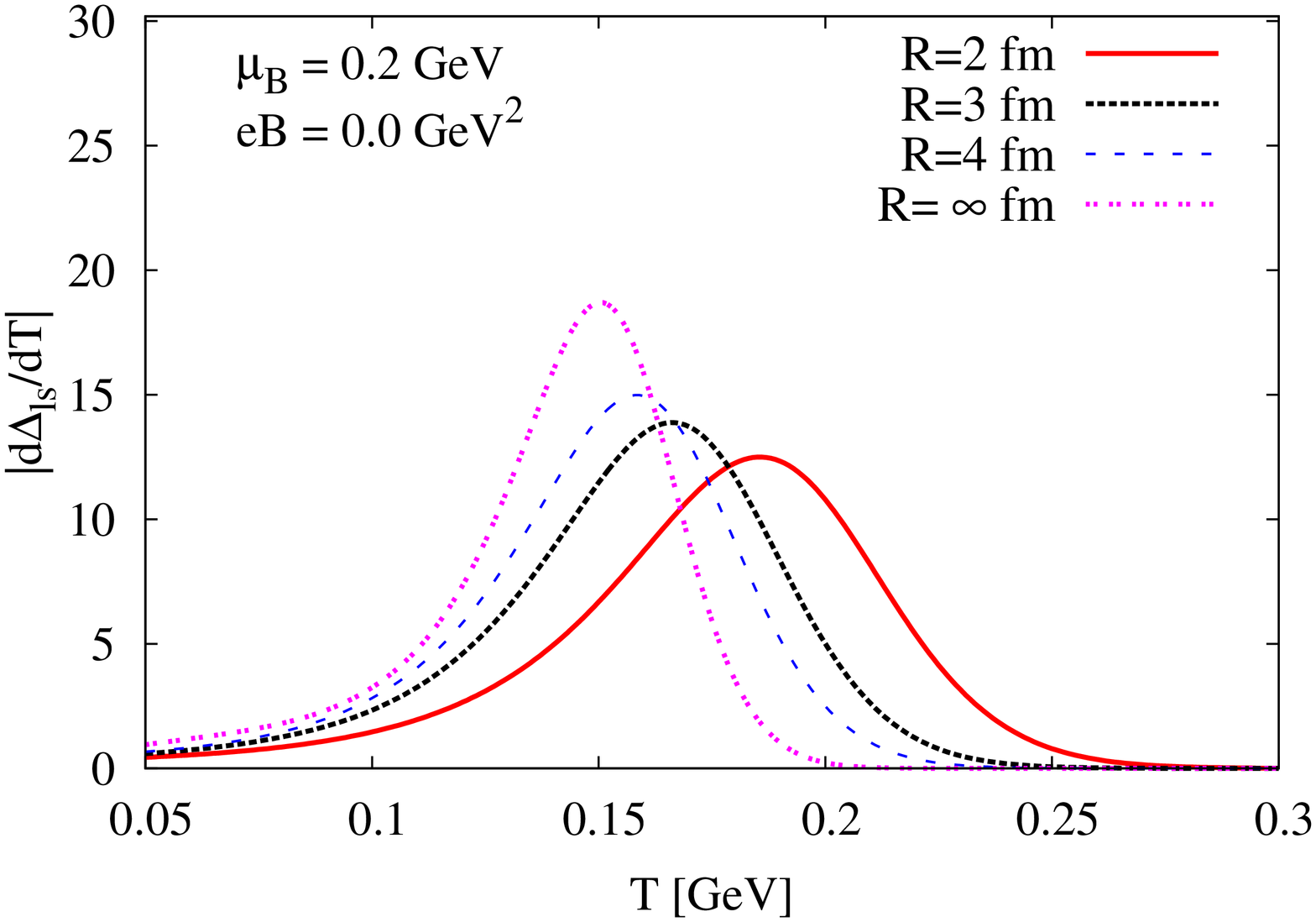}
\includegraphics[width=5.cm,angle=-90]{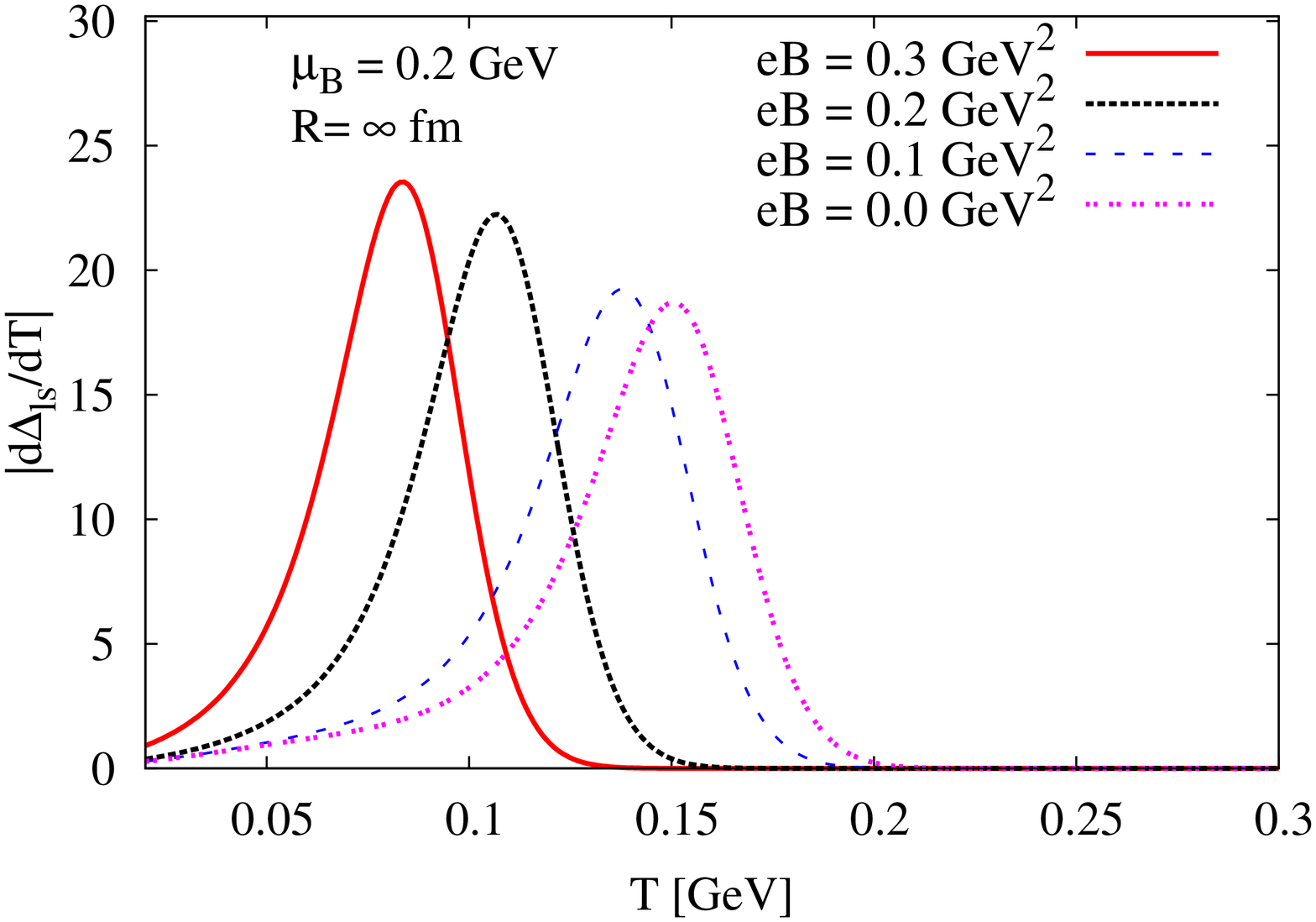}
\caption{(Color online) Same as in Fig.(\ref{fig:Delta}) but for $d\Delta_{l,s}/dT$.
 \label{fig:D_Delta}
 }
}
\end{figure*}
%
%
%
\begin{figure*}[htb]
\centering{
\includegraphics[width=5.cm,angle=-90]{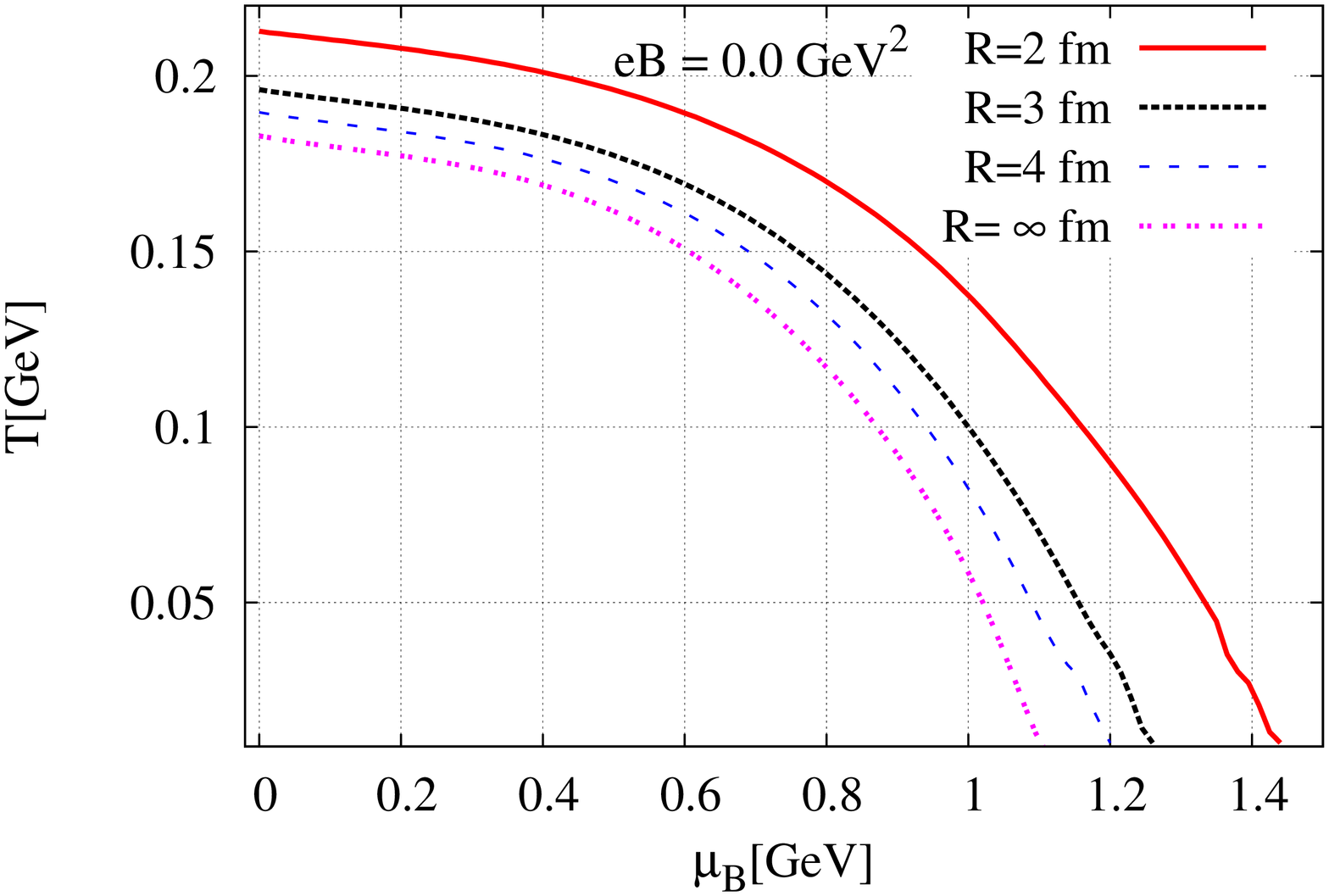}
\includegraphics[width=5.cm,angle=-90]{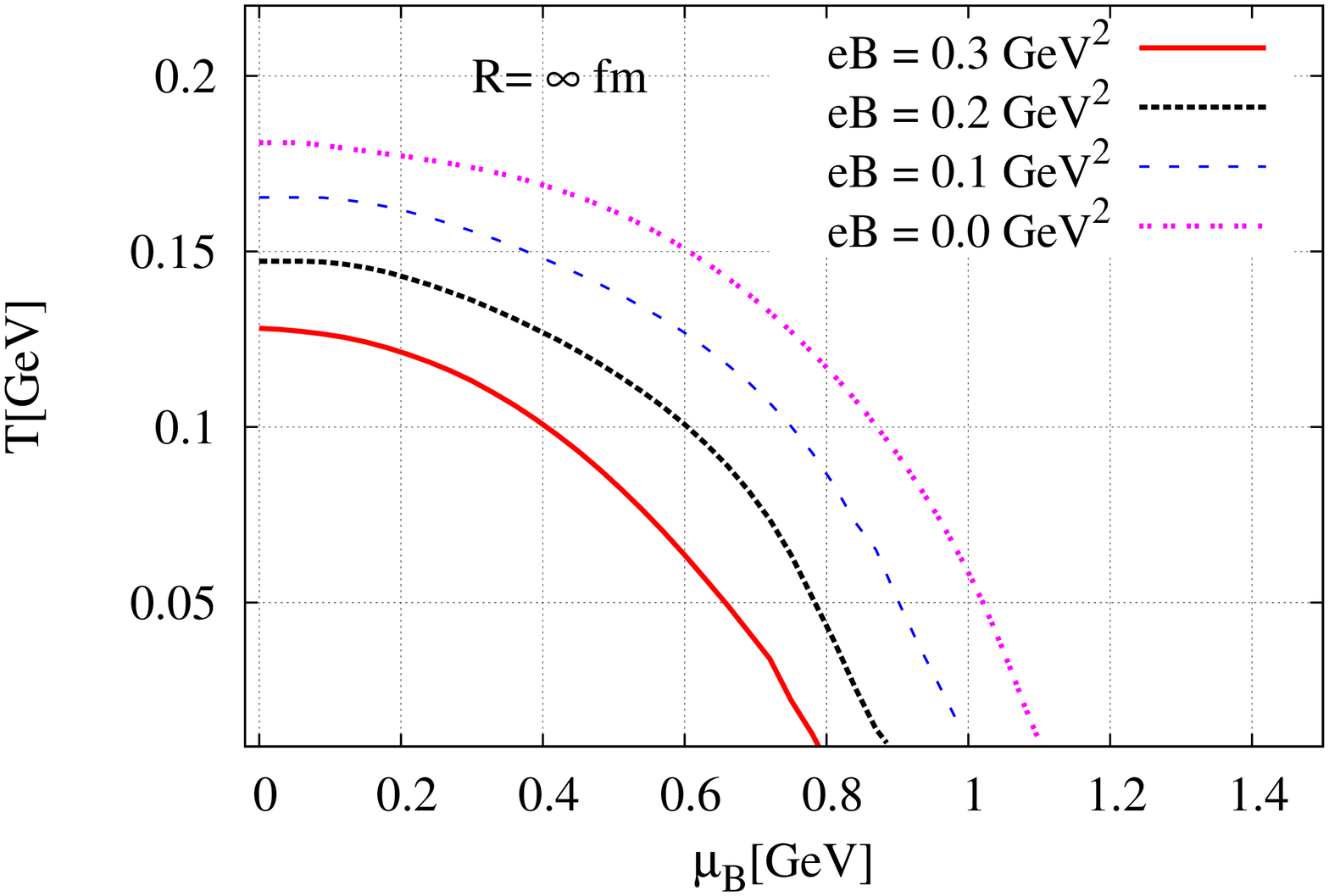}
\caption{(Color online) Chiral phase diagram for (i) different volumes selections for $eB = 0$ GeV$^2$
(left panel) and (ii) different  $eB$ selections for $R = \infty$~fm (right panel).
\label{fig:P-D}}
}
\end{figure*}

The pressure $P$ is easily obtained from the grand potential as, 
 \begin{eqnarray}
P &=& - \Omega(T, eB, \mu_f),  \label{Eq-Pr}
\end{eqnarray}
where Eq. (\ref{Eq-Pr}) expresses the [explicit] dependence of the pressure on the temperature, 
chemical potential, system volume and the magnetic field strength.  Coupled with the energy density $\epsilon$,
this pressure can also be used to obtain the trace anomaly $\Delta = \epsilon - 3 P$ and the 
equation of state $P/\epsilon$, and to study the influence of  finite volume and magnetic field effects on them.
Before discussing these effects, it is instructive to compare the values for $P$, $\Delta$ and $P/\epsilon$  obtained from 
our PLSM calculations (for $\mu_B = 0$ and $eB = 0$~GeV$^2$), to similar results from LQCD 
calculations~\cite{QCDL,QCD10}. Such comparisons  are shown in Fig.~\ref{fig:LQCD};  they indicate reasonable 
agreement between the PLSM and LQCD results for the model parameters summarized in Table~\ref{par_tab}.

Figure~\ref{fig:Pr} shows the temperature dependence of the normalized pressure
for different volume and magnetic field selections for two values of $\mu_B$. 
The left panels indicate an increase of the normalized pressure with volume which quickly trends towards 
the infinite volume value. The right panels show that the normalized pressure also increase with magnetic 
field strength, but do not trend toward a saturation value for the range of  magnetic field strengths studied.

Figure \ref{fig:Trec} shows the thermal behavior of the normalized trace anomaly for several 
volume and magnetic field selections for the previously used values of $\mu_B$. The left panels 
indicate that the normalized trace anomaly is insensitive to volume changes for $T \alt 0.2$~GeV.
For larger temperatures, the normalized trace anomaly decrease with increasing volume and 
quickly saturates to the infinite volume value. The right panels indicate a similar dependence 
of the normalized trace anomaly as a function of magnetic field strength for the full range of 
temperatures studied. That is, they show a decrease in magnitude with increasing magnetic field 
strength over the full temperature range.

The left panels of  Fig. \ref{fig:Cs2} show that $P/\epsilon$ is relatively insensitive to the volume at low 
temperatures. For higher temperatures, it shows an increase ($P/\epsilon$ gets softer) with volume which quickly saturates to the 
infinite volume value especially for the highest temperatures. The right panels of  Fig. \ref{fig:Cs2} also indicate an increase 
of $P/\epsilon$ with magnetic field strength, especially at low temperatures. Here, the magnitudes and trends are in stark 
contrast to those for the volume dependencies shown in the left panels.

\subsection{QCD phase diagram}
\label{subsec:IV}

The PLSM has two chiral order-parameters (the strange and non strange chiral condensates) which reflect the chiral phase transitions. To investigate finite volume and magnetic field effects on the $SU(3) 2+1$ PLSM chiral phase transition,  we use the normalized net-difference condensate $\Delta_{q,s}(T)$ as,
\begin{eqnarray}
 \Delta_ {q,s}(T) &=& \dfrac{\sigma_x - \dfrac{m_q}{m_s} \sigma_y}{\sigma_{x0} - \dfrac{m_q}{m_s} \sigma_{y0}},\label{Eq:Delta}
\end{eqnarray}
where $m_q$ ($m_s$) are non-strange (strange) quark masses. $\Delta_{q,s}(T)$ reflect the PLSM chiral phase transition.
Fig. \ref{fig:Delta} shows that the chiral phase transition is influenced by the effects of a finite volume and the magnetic field. The left panels indicate an increase in $\Delta_{q,s}(T)$ as the system volume is decreased. This trend contrasts with the influence of the magnetic field which results in a decrease of $\Delta_{q,s}(T)$ with increasing magnetic field strength. Thus, the effects of finite size have an opposing influence to those for the magnetic field.

The PLSM phase diagram (for a fixed volume and magnetic field strength) can be extracted with the 
aid of $\Delta_{q,s}(T)$. For fixed values of  $R$, $eB$ and $\mu_B$, the $d\Delta_{l,s}/dT$ is deduced as a function of temperature (cf. Fig. \ref{fig:D_Delta}). For the same baryon chemical potential, $d\Delta_{l,s}/dT$ will peak up at a characteristic point indicating the phase transition. Thus, the phase diagram can be generated by mapping such points for a broad range of baryon chemical potentials.
Fig. \ref{fig:P-D} illustrates the effects of finite volume and the magnetic field on the phase 
diagram.
The left panel shows that the phase boundary in the $(\mu_{B},T)$-plane of the PLSM phase-diagram, increases with decreasing system volume, i.e., both $T$ and $\mu_B$ increase as we decrease the system volume. 
A similar volume dependence has been reported Ref. \cite{Fraga_11}. The right 
panel shows that the effects of the magnetic field contrasts with the finite volume effects.
That is, the phase boundary shifts to lower values in the $(\mu_{B},T)$-plane as the 
magnetic field strength is increased.
%

\section{Conclusions}
\label{sec:conclusion}
In this work we have used the $2+1$ $SU(3)$ PLSM framework to investigate the properties of the 
QCD medium produced at finite volume and finite magnetic fields in heavy ion collisions. This 
model framework gives several thermodynamic quantities which compare well with those 
obtained in LQCD calculations for vanishing  $eB$  and $\mu_B$.
The PLSM calculations indicate that the confinement order parameters or Polyakov loops ($\phi$ and $\phi^*$) 
 are relatively insensitive to changes in the volume and the magnetic field strength. This contrasts with the  
chiral condensates which show a much larger sensitivity, albeit with much larger sensitivity for the non-strange
chiral order parameter. Both chiral condensates are found to increase with decreasing system volume, 
but decrease with increasing magnetic field strength. 

The PLSM calculations also indicate that several thermodynamic quantities ($P$, $\Delta$ and $P/\epsilon$) 
are significantly influenced by finite volume and finite magnetic field effects. Our combined study of 
PLSM thermodynamics and the chiral order parameters, suggests that the quark-hadron phase 
boundary is shifted to higher values of $\mu_{B}$ and $T$ with decreasing system volume,  
and to lower values of $\mu_{B}$ and $T$ with increasing magnetic field strength. Thus, 
the effect of a finite volume on the phase boundary is opposite to that for a finite magnetic field.
Additional studies geared at the influence of  a finite volume and a finite magnetic field on 
the location of the critical end point will be discussed in a future work.

\clearpage



%
\end{document}